\documentclass[twocolumn,showpacs,aps,prd,amssymb,nofootinbib,nobibnotes,floatfix,longbibliography]{aastex631}

\usepackage{graphicx}
\usepackage{dcolumn}
\usepackage{comment}
\usepackage{lipsum}
\usepackage{aas_macros}
\usepackage{bm}
\usepackage[dvipsnames]{xcolor}
\usepackage[percent]{overpic}
\definecolor{turquoise}{RGB}{96, 158, 160}
\definecolor{fuchsia}{RGB}{255, 0, 255}
\definecolor{darkorange}{RGB}{255, 127, 79}
\definecolor{red}{RGB}{220, 20, 60}
\definecolor{green}{RGB}{4, 235, 2}
\definecolor{darkyellow}{RGB}{255, 166, 6}

\usepackage{hyperref}
\usepackage[mathlines]{lineno}
\usepackage{amsmath}

\begin{document}

\preprint{APS/123-QED}

\title{Towards gravitational wave parameter inference for binaries with an eccentric companion}

\author{Kai Hendriks}
\thanks{Corresponding author: kai.hendriks@nbi.ku.dk}
\affiliation{Center of Gravity, The Niels Bohr Institute, Blegdamsvej 17, DK-2100, Copenhagen, Denmark}
\affiliation{Niels Bohr International Academy, The Niels Bohr Institute, Blegdamsvej 17, DK-2100, Copenhagen, Denmark}

\author{Lorenz Zwick}
\affiliation{Center of Gravity, The Niels Bohr Institute, Blegdamsvej 17, DK-2100, Copenhagen, Denmark}
\affiliation{Niels Bohr International Academy, The Niels Bohr Institute, Blegdamsvej 17, DK-2100, Copenhagen, Denmark}

\author{Pankaj Saini}
\affiliation{Center of Gravity, The Niels Bohr Institute, Blegdamsvej 17, DK-2100, Copenhagen, Denmark}
\affiliation{Niels Bohr International Academy, The Niels Bohr Institute, Blegdamsvej 17, DK-2100, Copenhagen, Denmark}

\author{János Takátsy}
\affiliation{Institut f\"ur Physik und Astronomie, Universit\"at Potsdam,
Haus 28, Karl-Liebknecht-Str. 24-25, Potsdam, Germany}
\affiliation{Niels Bohr International Academy, The Niels Bohr Institute, Blegdamsvej 17, DK-2100, Copenhagen, Denmark}

\author{Johan Samsing}
\affiliation{Center of Gravity, The Niels Bohr Institute, Blegdamsvej 17, DK-2100, Copenhagen, Denmark}
\affiliation{Niels Bohr International Academy, The Niels Bohr Institute, Blegdamsvej 17, DK-2100, Copenhagen, Denmark}

\date{\today}

\begin{abstract}
We introduce a complete model for dephasing due to  line-of-sight acceleration (LOSA) in gravitational wave (GW) signals from stellar-mass binary black holes (BBHs) in three-body systems. Our prescription provides curvature- and projection-dependent phase features that are not recovered by local-expansion-based treatments. We perform parameter-space surveys and mock parameter inferences assuming the nominal sensitivity of the Einstein Telescope (ET) to identify the regime where the time-varying LOSA allows for separate constraints on the outer orbital parameters, in particular the tertiary mass and distance. We estimate that ET may detect a few to tens of such systems per year, provided that all binaries merge dynamically, and demonstrate that these constraints can be used to directly discriminate between a dynamical and AGN origin for BBHs. Finally, we reanalyse the GW190814 event and four O4a events finding no evidence for LOSA, with the previously claimed LOSA in GW190814 disappearing when a sufficiently long data segment is used.
\end{abstract}

\section{Introduction}
Understanding the formation pathways of binary black holes (BBHs) remains a central challenge in gravitational-wave (GW) astrophysics, particularly following the now $\sim$200 BBH detections reported by the LIGO-Virgo-KAGRA (LVK) Collaboration \citep{gwtc-4}. Several formation channels have been proposed, including
isolated binary stars \citep{2012ApJ...759...52D, 2013ApJ...779...72D, 2015ApJ...806..263D, 2016ApJ...819..108B,
2016Natur.534..512B, 2017ApJ...836...39S, 2017ApJ...845..173M, 2018ApJ...863....7R, 2018ApJ...862L...3S, 2023MNRAS.524..426I},
dynamical interactions within dense stellar clusters \citep{2000ApJ...528L..17P, Lee:2010in,
2010MNRAS.402..371B, 2013MNRAS.435.1358T, 2014MNRAS.440.2714B,
2015PhRvL.115e1101R, 2015ApJ...802L..22R, 2016PhRvD..93h4029R, 2016ApJ...824L...8R,
2016ApJ...824L...8R, 2017MNRAS.464L..36A, 2017MNRAS.469.4665P, samsing2018a, 2018MNRAS.tmp.2223S, 2019arXiv190711231S, 2021MNRAS.504..910T, 2022MNRAS.511.1362T},
galactic nuclei (GN) \citep{2009MNRAS.395.2127O, 2015MNRAS.448..754H,
2016ApJ...828...77V, 2016ApJ...831..187A, 2016MNRAS.460.3494S, hoa18, 2018ApJ...865....2H,2019ApJ...885..135T, 2019ApJ...883L...7L,2021MNRAS.502.2049L, 2023MNRAS.523.4227A},
active galactic nuclei (AGN) discs \citep{2017ApJ...835..165B,  2017MNRAS.464..946S, 2017arXiv170207818M, 2020ApJ...898...25T, 2022Natur.603..237S, 2023arXiv231213281T, Fabj24},
single-single GW captures of primordial black holes \citep{2016PhRvL.116t1301B, 2016PhRvD..94h4013C,
2016PhRvL.117f1101S, 2016PhRvD..94h3504C},
and very massive stellar mergers \citep{Loeb:2016, Woosley:2016, Janiuk+2017, DOrazioLoeb:2018}. If contributing to the overall merger rate, each channel is expected to produce distinct populations of mergers with characteristic properties in specific observables such as the binary eccentricity \citep[e.g.][]{2006ApJ...640..156G, 2014ApJ...784...71S, 2017ApJ...840L..14S, Samsing18a, Samsing2018, 2018ApJ...855..124S,
2018MNRAS.tmp.2223S, 2018PhRvD..98l3005R, 2019ApJ...881...41L,2019ApJ...871...91Z, 2019PhRvD.100d3010S, 2019arXiv190711231S}, the relative spin orientation of the
merging BBHs \citep[e.g.][]{2000ApJ...541..319K, 2016ApJ...832L...2R,2018ApJ...863...68L}, or the mass
distribution \citep[e.g.][]{2017ApJ...846...82Z,2021MNRAS.505.3681S}.

Instead of focusing solely on population-level characteristics, recent studies have proposed investigating environmental effects (EEs) as an additional means to constrain BBH formation scenarios. EEs are modifications to a vacuum waveform caused by astrophysical perturbations to the GW source \citep[e.g.][]{1993chakrabarti,1995ryan,2008barausse,2007levin,kocsis,2014barausse,inayoshi2017,2017meiron,2017Bonetti,2019alejandro,2019randall,2020cardoso,DOrazioGWLens:2020,2022liu,2022xuan,garg2022,2022cole,2022chandramouli,2022sberna,2023zwick,2023Tiede,2024dyson,2022destounis,2022cardoso,2020caputo,2024zwicknovel,Derdzinksi:2021,2024basu,2024santoro,2007levin,Derdzinksi:2021,garg2022,2022speri,2024duque,liu2015,2021toubiana,2023zwick,2024barandiaran,2024zwick, 2025PhRvL.134h1402S}. Over the last few years, several works have highlighted how EEs may be used as a powerful tool, enabling to identify \textit{smoking gun signatures} of binary formation channels on an individual-event basis. Compact object binary formation pathways that are characterised by strong interactions with the environment, such as the dynamical \citep{2000ApJ...528L..17P, Lee:2010in,
2010MNRAS.402..371B, 2013MNRAS.435.1358T, 2014MNRAS.440.2714B,
2015PhRvL.115e1101R, 2015ApJ...802L..22R, 2016PhRvD..93h4029R, 2016ApJ...824L...8R,
2016ApJ...824L...8R, 2017MNRAS.464L..36A, 2017MNRAS.469.4665P, 2018MNRAS.tmp.2223S, 2020PhRvD.101l3010S, 2021MNRAS.504..910T, 2013ApJ...773..187N, 2014ApJ...785..116L, 2016ApJ...816...65A, 2016MNRAS.456.4219A, 2017ApJ...836...39S, 2018ApJ...864..134R, 2019ApJ...883...23H, 2021MNRAS.502.2049L, 2022MNRAS.511.1362T, 2025PhRvD.112j3047S} and AGN \citep{2017ApJ...835..165B,  2017MNRAS.464..946S, 2017arXiv170207818M, 2020ApJ...898...25T, 2022Natur.603..237S,
2023arXiv231213281T, trani2024, Fabj24} channels, are expected to produce a significant fraction of sources with detectable EE for next generation ground-based detectors \citep{2025zwick}.

In particular, the most promising avenue is the detection of Rømer delays, or line-of-sight acceleration (LOSA)\footnote{The terms line-of-sight/centre-of-mass acceleration and Rømer delay phase shift (or dephasing) refer to exactly the same type of EE. We use different naming conventions depending on the context.} due to a nearby third body \citep{2017meiron,bonvin2017,Vijaykumar2023-qg,2024samsing,kai2024,kai22024,2018PhRvD..98f4012R}. Rømer delays induce a frequency-dependent shift $\delta \psi_{\rm R}$ in the GW phase with an amplitude proportional to the gravitational acceleration and in its simplest form (a constant LOSA) a characteristic frequency scaling of $f^{-13/3}$:
\begin{align}
    \label{eq:deph}
    \delta \psi_R \propto \frac{m_3}{R_3^2} \, f^{-13/3},
\end{align}
where $m_3$ is the third body mass, $R_3$ the distance between the third body and the binary centre of mass, and $f$ is the detector frame GW frequency. While the detection of dephasing with this frequency scaling would provide incontrovertible evidence for the presence of a third body, the size of the dephasing itself is determined by a degenerate combination of third body mass and distance. Therefore, it would be impossible to precisely determine the properties of the third body and distinguish between perturbations from another stellar-mass object (dynamical channel) or a supermassive BH (AGN channel). Previous studies have improved upon constant-acceleration models by incorporating higher-order time derivatives of the LOSA which partially capture the curvature of the outer orbit and yield more accurate dephasings in certain regimes \citep{2024MNRAS.527.8586T, 2025PhRvD.112h4034T, 2025arXiv250622272T}. However, these treatments rely on local Taylor expansions and therefore cannot reproduce the full, strongly time-dependent structure of the acceleration that arises in eccentric Keplerian orbits or from evolving projection geometry relevant for dynamical interactions in dense stellar clusters.

In this work we introduce a model of Rømer delay which includes arbitrary outer orbit eccentricity and incorporates time-dependent projection effects. This global treatment reveals features in the phase evolution that cannot be captured by expansion-based approaches. We show explicitly in what astrophysical scenarios the mass–distance degeneracy can be broken. Additionally, motivated by the potential for degeneracy-breaking, we re-analyse GW190814, which was previously reported to exhibit a measurable LOSA \citep{2024Han}, as well as analyse several promising events from the LVK O4a observing run \citep{gwtc-4}.

This paper is structured as follows: in Sec. \ref{sec:methods}, we lay out the waveform and dephasing models used in this work, as well as the mathematical frameworks for the parameter inference. In Sec. \ref{sec:exploration}, we explore the parameter space of the dephasing model and investigate the interplay between different outer orbital parameters. Subsequently, in Sec. \ref{sec:mock_inference}, we generate several mock GW events that include our updated dephasing model and perform parameter inference assuming ET sensitivity. Lastly, in Sec. \ref{sec:lvk_inference}, we show the results from our LOSA inference of GW190814 and several O4a events. \\ \\



\section{Methods}
\label{sec:methods}
\begin{figure}
    \centering
    \includegraphics[width=0.9\linewidth]{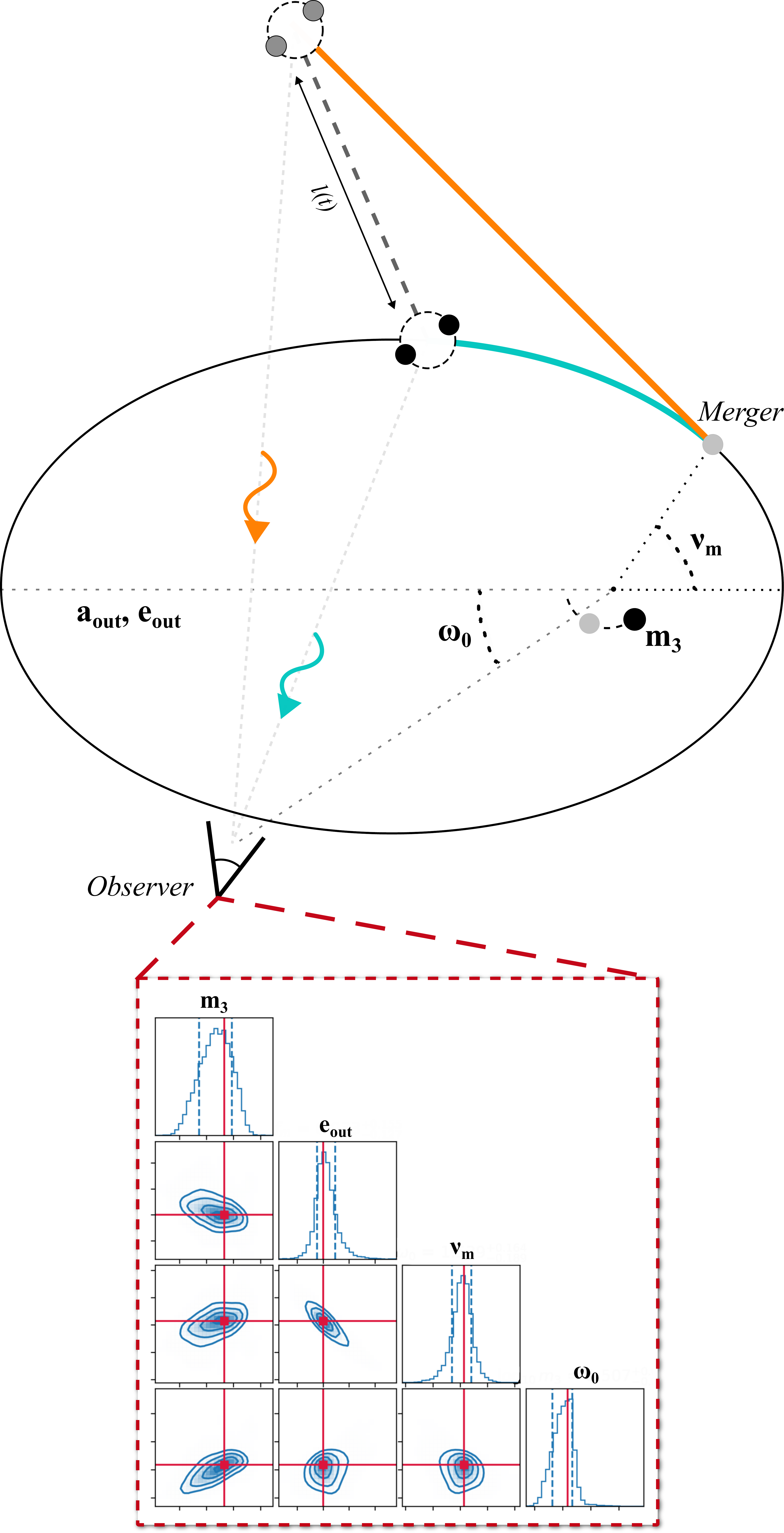}
    \caption{Schematic depiction of the premise of this work, in which we study the detectability of the GW phase shift in signals from BBHs whose COM is on an orbit around a third BH. By modelling the outer eccentricity and projection effects, one can in principle constrain the properties of the three-body system.}
    \label{fig:setup}
\end{figure}
\subsection{Waveforms and SNR calculations}
\label{sec:waveforms}
While it is known that BBHs in realistic dynamical environments can in fact retain a small but non-negligible eccentricity upon entering the detector band \citep[e.g.][]{samsing2018a, 2020shaw}, we assume a circular inner orbit. The circular assumption allows for fast waveform evaluation and for isolating the impact of the environmental dephasing itself. Note however, that eccentricity introduces additional harmonics that encode information about the earlier stages of the inspiral, which can significantly enhance the detectability of environmental effects \citep{pedpaper,2025arXiv251104540Z}.

We employ several classes of circular waveform models. For our parameter space survey in Sec. \ref{sec:exploration} and mock inference in Sec. \ref{sec:mock_inference}, we use Fourier space Newtonian waveforms in the stationary phase approximation \citep{1994cutler,blanchet2014}: 
\begin{align}
\label{eq:GWnew}
    \tilde{h}(f) = \sqrt{\frac{5}{24}} \pi^{-2/3} \frac{Q}{D(z)}\frac{(G\mathcal{M}_z)}{c^{3/2}}^{5/6}f^{-7/6} \exp \left[ i \psi_{\rm vac} \right],
\end{align}
where $f$ is the observer frame GW frequency, $D(z)$ is the luminosity distance, $\mathcal{M}_z$ is the red-shifted chirp mass, and $Q$ a geometric pre-factor. We assume an all-sky average with $Q=2/5$ and additionally multiply with 3/2 to account for the geometry of the detector \citep{2012PhRvD..86l2001R}. The GW phase $\psi_{\rm vac}$ is given by:
\begin{align}
    \psi_{\rm vac} &= 2\pi f t_c - \phi_c - \frac{\pi}{4}-  \frac{3}{128}\left(\frac{\pi G \mathcal{M}_z f}{c^3} \right)^{-5/3}.
    \label{eq:gw-phase}
\end{align}
Here we set the constant phase offset $\phi_{\rm c}$ and integration constant $t_{\rm c}$ representing the phase and time at coalescence both to 0, as they do not influence simple estimates of the BBH SNR (though they are crucial for full parameter inference). The vacuum phase (Eq. \ref{eq:gw-phase}) can be supplied by additional dephasing prescriptions as a simple model for EE, as is done in many works \citep[e.g.][]{2025zwick, 2024samsing, kai22024, 2024lensingsamsing, zwicklensing, pedpaper}. Then:
\begin{align}
    \psi_{\rm tot} = \psi_{\rm vac} + \psi_{\rm EE}.
\end{align}

For the analysis of GW190814 and the O4a events in Sec. \ref{sec:lvk_inference}, we employ the \texttt{IMRPhenomXPHM} waveform, enabling a direct comparison with \citet{2024Han} and the GWTC-4 catalogue \citep{gwtc-4}. For consistency with GWTC-4, we additionally include the \texttt{SpinTaylor} extension for the O4a detections.

The SNR of a waveform $h$ is given by the square root of the noise weighted inner product $<h,h>$, where:
\begin{align}
    \label{eq:innerprod}
    \left< h_1 , h_2\right> = 2\int_0^{\infty} \frac{\tilde{h}_1\tilde{h}_2^{*} + \tilde{h}_1^{*}\tilde{h}_2}{S_{n}(f')}{\rm{d}}f'
\end{align}
where $\tilde{h}_i$ is the Fourier transform of  $h_i$, whereas $S_{n}$ characterises the noise profile of the given GW detector. Where this approach is in principle independent of the choice of GW detector, we solely focus on ET \citep{2020maggioreet, 2025abac_et} sensitivity in this work as EEs are expected to only start to become detectable in next-generation GW observatories \citep{2025zwick}.

A simple estimate for the detectability of an EE is the so-called $\delta$SNR criterion:
\begin{align}
\label{eq:dSNRcrit}
   \delta \text{SNR}^2 &\equiv \left<\Delta h,\Delta h \right> > \mathcal{C}^2, \\
   \Delta h&= h_{\rm vac} - h_{\rm EE}.
\end{align}
Here $h_{\rm vac}$ represents the vacuum waveform, while $h_{\rm EE}$ the perturbed waveform with the addition of EE. The threshold $\mathcal{C}$ is commonly chosen to be $\mathcal{O}(10)$ to assure sufficient mismatch. We note that we will supply our parameter space exploration using Eq. \ref{eq:dSNRcrit} with several Bayesian parameter inference experiments, as detailed in Sec. \ref{sec:mock_inference}.

In this work, we will also investigate the prospects for distinguishing between different types of EE within the same waveform. Therefore, we introduce the additional concept of a residual SNR between two waveforms enhanced with different EE prescriptions (EE1 and EE2). We dub the concept of "delta--$\delta$SNR", defined as:
\begin{align}
\label{eq:deltadSNR}
   \Delta\delta \text{SNR}^2 &\equiv \left<\Delta h,\Delta h \right> > \mathcal{C}^2, \\
   \Delta h&= h_{\rm EE1} - h_{\rm EE2}.
\end{align}
This serves as a useful tool to survey large portions of parameter space.

In our application, the two EEs correspond to the following choices. The waveform $h_{\mathrm{EE1}}$ incorporates the eccentric dephasing described in Eq.~\ref{eq:dephasing}. The comparison waveform $h_{\mathrm{EE2}}$ instead includes a generic dephasing term of the form
\begin{align}
    d\psi = A f^{-13/3},
    \label{eq:deph_LOSA_gen}
\end{align}
which captures the leading frequency dependence expected from a LOSA. For each realisation of $h_{\mathrm{EE1}}$, we construct $h_{\mathrm{EE2}}$ such that both waveforms exhibit the same dephasing amplitude when the binary enters the ET frequency band ($f_{\mathrm{in}}\sim 2\,\mathrm{Hz}$ for our example systems). The constant $A$ is determined by matching $d\psi_{\mathrm{EE2}}$ to $d\psi_{\mathrm{EE1}}$ at this frequency. This procedure provides the "best-fit" non-eccentric dephasing with the same initial magnitude, thereby allowing the resulting $\Delta\delta \mathrm{SNR}$ to quantify how distinguishable the true eccentric signal is from its closest generic (non-eccentric) counterpart.

\subsection{Eccentric Rømer delay dephasing}
\label{sec:dephasing}
The effect of dephasing due to Rømer delay has been studied in detail in multiple works \citep[e.g.][]{2011PhRvD..83d4030Y,2018PhRvD..98f4012R, 2024samsing, kai22024, Vijaykumar2023-qg, 2024MNRAS.527.8586T}. Central to this paper is our previous work in \cite{kai22024}, where we investigated the Rømer dephasing allowing for \textit{both} the binary orbit and the outer orbit to be eccentric. We note two key findings from this previous work: \textit{i)} an eccentric outer orbit can increase the amount of dephasing with several orders of magnitude compared to its circular equivalent, and \textit{ii)} the outer eccentricity can have clear distinguishable features in the dephasing. These features make the dephasing deviate from the power law in Eq. \ref{eq:deph}.

A first quantification of dephasing from 3-body interactions in realistic globular-cluster-like settings was carried out in \cite{kai2024}, showing that specific configurations exist that generate exceptionally large dephasing. A more sophisticated parameter study, carried out in this work, which incorporates eccentricity effects and predicts in more detail the detectability of Rømer dephasing is therefore a logical and necessary next step.

To describe the Rømer dephasing for an eccentric outer orbit, we employ the framework described in \cite{kai22024}. We define\footnote{Note that in our previous work we defined $d\psi = \frac{2\pi}{c} l(t) f_\mathrm{orb}$, to make the transition between circular and eccentric binaries smoother. Here we choose the definition in Eq. \ref{eq:dphi_general}, which is the appropriate form for application to full waveform models.} the Rømer dephasing as
\begin{align}
    d\psi = \frac{2\pi}{c} l(t) f,
    \label{eq:dphi_general}
\end{align}
where $l(t)/c$ is the difference in light travel time between the observed binary whose COM is on a curved path (turquoise in Fig. \ref{fig:setup}) and a fictional reference binary with a constant speed (orange in Fig. \ref{fig:setup}). When the outer orbit is circular, the Rømer time delay $l(t)/c$ has a closed analytical form \citep{2017meiron, 2024samsing}. In the outer eccentric case, without approximating the local curvature (e.g. via a Taylor expansion), the Rømer delay can only be obtained numerically as its evaluation requires solving the Kepler equation. We find $l(t)/c$ numerically and express it as $l(f)/c$ in terms of the GW frequency $f$ to be able to apply it to a Fourier domain waveform. As mentioned before, we assume that the inspiralling binary is circular, which yields for the inspiral time $t$ \citep{peters1964,2009yunes}:
\begin{align}
    t_\text{circ} = \frac{5 \mathcal{M}}{256} G^{-5/3}c^5(2\pi\mathcal{M}\frac{f}{2})^{-8/3},
\end{align}
where $\mathcal{M}$ is the chirp mass, and $f$ the GW frequency. Similar to Eq. \ref{eq:gw-phase}, we here ignore the constant offset $t_c$ which is not important for our SNR estimates.

We also introduce the quantity $F(\omega_0)$, which is a projection factor that depends on the angle $\omega_0$ between the observer and the major axis of the outer ellipse. More explicitly, $F(\omega_0,f) = \cos(\omega_0 + \alpha(t))$ where $\alpha(t)$ is the angle between the $x-$ and $y-$component of the $\textbf{l}(t)$ vector before the $\omega_0$ rotation is applied to the system. The final form of the Rømer dephasing is then
\begin{align}
    d\psi_R(f) = F(\omega_0) \frac{2\pi}{c} l(f) f,
    \label{eq:dephasing}
\end{align}
which therefore accounts for both eccentricity and projection modifications to the commonly used constant acceleration dephasing prescription.

In this study we assume that the observer lies in the plane of the outer orbit, i.e. $\sin\iota = 1$, where $\iota$ is the inclination angle. This idealisation allows the environmental dephasing to be isolated and examined in a controlled manner. It is important to note that this means that the recovered parameters should be treated as lower bounds.

In this work we adopt a sign convention for the true anomaly at merger, $\nu_m$, such that the example in Fig. \ref{fig:setup} is defined to be positive. Consequently, a negative value (for example, $\nu_m=-\pi/4$) indicates that the binary passes the outer pericentre during its inspiral (given a sufficiently long observation time). The viewing-angle parameter $\omega_0$ in Fig. \ref{fig:setup} is shown with a negative sign under this convention. A positive value corresponds to the mirror configuration across the semi-major axis of the outer orbit.

We apply the Rømer dephasing externally to the waveform, as follows \citep{pedpaper}:
\begin{align}
    \tilde{h}_\text{pert} (f) = \tilde{h}_\mathrm{vac} (f) \exp[i(-d\psi_R(f))].
    \label{eq:hpert}
\end{align}

In Sec. \ref{sec:lvk_inference}, we customise the \texttt{bilby} waveform by applying a leading-order LOSA dephasing to it according to \cite{bonvin2017}:
\begin{align}
    d\psi(f) = \frac{25}{65536\eta^2} \left(\frac{GM}{c^3}\right) \left(\frac{a}{c}\right) \left(\frac{GM}{c^3}f\right)^{-13/3},
    \label{eq:dephasing_LOSA}
\end{align}
where $a$ is the LOSA, $M$ the total mass, and $\eta$ is the symmetric mass ratio. We apply it to the waveform in the same fashion as Eq. \ref{eq:hpert}.

\subsection{MCMC parameter inference}
\label{sec:mcmc}
In this work, we employ two methodologies to perform parameter inference experiments.
For the tests performed in Sec. \ref{sec:mock_inference}, we adopt the Newtonian waveforms described in section \ref{sec:waveforms}, supplied by a dephasing prescription for EE. We assume that the binary coalescence time $t_{c}$ is fixed by a measurement of the merger and ring-down portions of the GW signal, and truncate the waveform when the binary reaches the innermost stable circular orbit. To perform the parameter inference we use Monte-Carlo-Markhov-Chain  (MCMC) methods. We employ the convenient sampler \texttt{emcee} \citep{2013emcee}, typically running 18 parallel walkers for 10,000 steps.
The likelihood $\mathcal{L}$ for the MCMC tests is given by:
\begin{align}
    \mathcal{L}\left(\Theta \right) \propto \exp\left[ - \big<h(\Theta) - h(\Theta_{\rm GT}),h(\Theta) - h(\Theta_{\rm GT}) \big> \right]
\end{align}
where $\Theta_{\rm GT}$ is a vector of ``ground-truth" parameters for the injected signal. The noise is stationary and determined by the power spectral density of the given GW detector, which is ET in this work \citep{2020maggioreet, 2025abac_et}. To facilitate convergence, we initialise the walkers near the injected true values. This approach also eliminates the need to define priors for the vacuum binary parameters (which is only suitable for high SNR signals. These choices for the initialisation set-up are made purely for convenience and have no impact on the final results unless stated otherwise.

For the parameter inference of the real GW data (Sec. \ref{sec:lvk_inference}), we use the open-source code \texttt{bilby} \citep{bilby_paper}. This package is also MCMC-based, and has the convenient functionality of easy access to raw data of LVK events and waveforms. For all analyses, our sampler settings are \texttt{sampler='dynesty'}, \texttt{nlive=1000}, \texttt{naccept=60}, \texttt{sampler='acceptance-walk'}, and \texttt{dlogz=0.1}.

\section{Exploration of parameter space}
\label{sec:exploration}
\begin{figure*}
    \centering
    \includegraphics[width=1\linewidth]{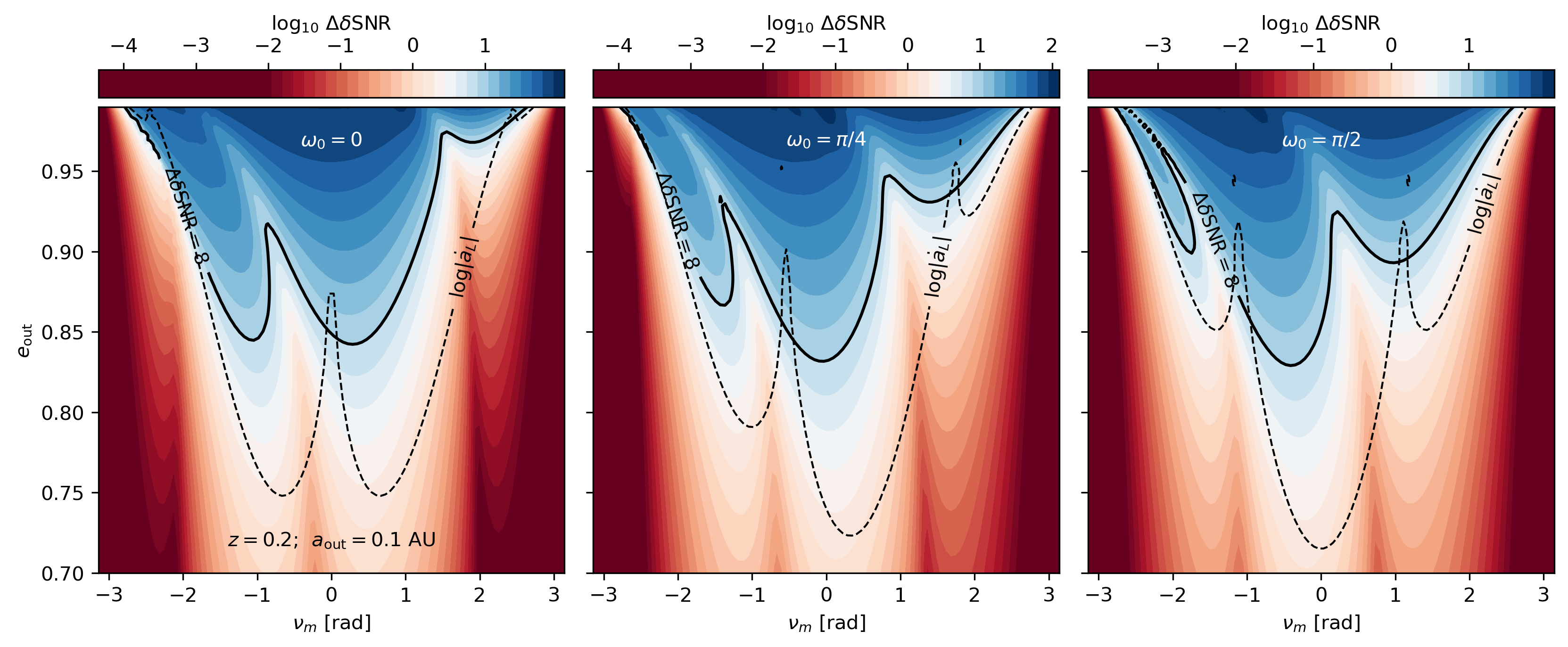}
    
    \caption{$\Delta \delta$SNR contours as a function of outer orbit eccentricity $e_{\rm out}$ and merger true anomaly $\nu_m$, for $m_1 = m_2 = m_3 = 8 $ M$_{\odot}$. The panels show the results for three different choices of $\omega_0$, i.e. three different viewing angles. Typically, the largest residuals occur for mergers just before of just after pericentre, where both acceleration and eccentricity signatures are maximal. The black solid contour represents configurations with $\Delta\delta\mathrm{SNR}=8$, which is approximately where it becomes distinguishable from circular dephasing. Additionally, we overplot a contour of constant LOSJ (dashed), according to Eq. \ref{eq:losj}.}
    \label{fig:deltadelta}
\end{figure*}
Before performing parameter estimation tests, we first survey the relevant regions of parameter space. Our goal is to identify where dephasing induced by a constant LOSA can be distinguished from dephasing produced by a tertiary companion on an eccentric outer orbit. To this end, we employ the $\Delta\delta\mathrm{SNR}$ criterion introduced in Sec. \ref{sec:methods}. This metric does not, by itself, guarantee that the two effects are separable in a full Bayesian analysis. Rather, it quantifies whether the residual between the two waveform models carries sufficient power to be observationally detectable. A subsequent Bayesian inference step (Sec. \ref{sec:mock_inference}) is then required to determine whether distinguishability persists once parameter degeneracies are taken into account.

Our methodology is as follows: We consider a representative, high-SNR binary with component masses $m_1 = m_2 = 8,M_\odot$, perturbed by a tertiary companion of mass $m_3 = 10,M_\odot$, and placed at redshift $z = 0.2$. For this system, we compute $\Delta\delta\mathrm{SNR}$ values for waveforms that include a Rømer delay modeled either with the full eccentric prescription (Eq. \ref{eq:dephasing}) or with the best-fit approximation of the circular version (Eq. \ref{eq:deph_LOSA_gen}), as described in Sec. \ref{sec:dephasing}. The resulting $\Delta\delta\mathrm{SNR}$ quantifies the residual power between the waveform perturbed by an eccentric outer orbit and that obtained using a constant LOSA.

Our choice of fiducial parameters is motivated by astrophysical considerations and by the need to obtain signals with measurable dephasing. The binary component masses $m_1$ and $m_2$ are selected to lie near the peak of the LVK mass distribution, making them representative of the underlying population \citep{gwtc-4, gwtc4-population}. These relatively low masses also maximise the accumulated dephasing, which is advantageous for the present study. In dense stellar clusters, three-body encounters are most efficient for nearly equal mass objects \citep[e.g.][]{cmc_paper, cmcpaper, 2015ApJ...800....9M}, and for this reason we choose a tertiary mass $m_{3}$ that is comparable to the binary masses, both in this survey and in the mock inference presented below. The adopted redshift of $z = 0.2$ yields a high SNR for ET, which is required for the dephasing effects to be detectable. This redshift lies well within the expected horizon of ET \citep{2020maggioreet}, and although it does not correspond to the "typical" detection distance, systems at this redshift are still expected to contribute merger rates of several tens per $\mathrm{Gpc}^{3}$ per year \citep{gwtc4-population}. Finally, our fiducial choice of $a_{\mathrm{out}} = 0.1$ AU is typical of cluster environments \citep{antonini2016}. In the sections below we also explore values as small as $a_{\mathrm{out}} \sim 0.01$ AU, which remain within the range expected in the densest regions of clusters and are therefore astrophysically well motivated \citep{2021ApJ...923..126S}.

Figure~\ref{fig:deltadelta} collects some representative results. The panels show $\Delta\delta\mathrm{SNR}$ contours as a function of the outer-orbit eccentricity and the true anomaly at merger, $\nu_m$, for three choices of the projection angle $\omega_0$, representing here three different viewing angle configurations. Here, low-eccentricity tertiary orbits are effectively indistinguishable from circular ones. Instead, detectable differences emerge in certain regions of high eccentricity, with the precise threshold modulated by $\nu_m$. The largest residuals occur when the merger takes place in a region with strong variation in the LOSA over an inspiral timescale. Typically, this occurs just before or just after pericentre passage, a configuration in which both the instantaneous acceleration and the discrepancy between the eccentric and constant-acceleration dephasing prescriptions are maximised. 

In our setup, the measured LOSA expresses itself analytically as
\begin{equation}
    a_\mathrm{L}(t) = F(\omega_0, \nu(t)) \times Gm_3 \frac{(1 + e_\text{out}\cos\nu(t))^2}{p_\text{out}^2} \:,
    \label{eq:losa-ecc}
\end{equation}
where we have used the fact that the orbital radius is $R(\nu) = a_\mathrm{out} (1-e_\mathrm{out}^2)/(1+e_\mathrm{out}\cos\nu) = p_\mathrm{out}/(1+e_\mathrm{out}\cos\nu)$, with $p_\mathrm{out}$ the semi-latus rectum. The combined time dependence of the separation $R$ (from the eccentricity) and the multiplicative factor $F(\omega_0, \nu(t))$ (from the projection effects) determines the structure of the measured LOSA. In the limit where the change in LOSA over time is relatively constant, we can write $F(\omega_0, \nu(t)) \approx \cos(\nu + \omega_0)$. The line-of-sight jerk (LOSJ) then becomes
\begin{align}
\begin{split}
|\dot{a}_L| \approx \frac{G m_3}{R^2}\dot{\nu} &\Bigg[ \frac{2e_\mathrm{out}}{1+e_\mathrm{out}\cos\nu}  \sin\nu \cos(\nu+\omega_0) \\ 
    &\quad + \sin(\nu + \omega_0) \Bigg],    
\end{split}
\label{eq:losj}
\end{align}
with $\dot{\nu} = (G m_3 p_\mathrm{out})^{1/2}/R^2$ from angular momentum conservation. We overlay a contour of constant LOSJ in Fig. \ref{fig:deltadelta} (evaluated at $\nu=\nu_m$). The imprint of the LOSJ is clearly reflected in the overall morphology of the $\Delta\delta\mathrm{SNR}$ contours, most notably in the locations of the $\Delta\delta\mathrm{SNR}$ peaks. In general, and especially at the highest eccentricities, the constant–LOSJ approximation of Eq. \ref{eq:losj} does not perfectly reproduce the shape of the $\Delta\delta\mathrm{SNR}$ contours. The contours become visibly skewed relative to the LOSJ prediction, indicating that higher-order derivative of the LOSA contribute appreciably in this regime and that the full model is required for an accurate description.

\begin{figure}
    \centering
    \includegraphics[width=1\linewidth]{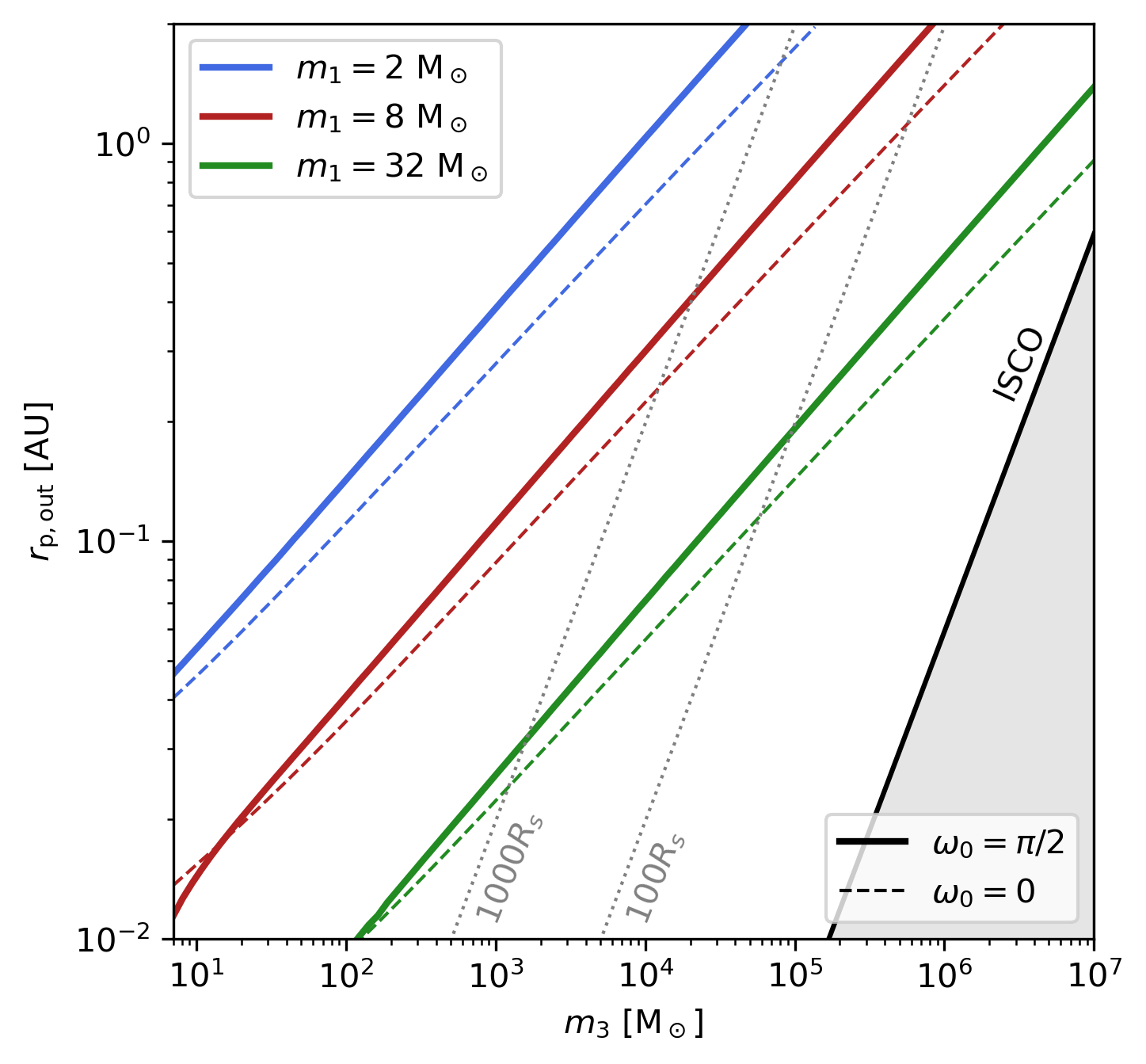}
    \caption{Tertiary pericentre distance required to achieve $\Delta \delta{\rm SNR} = 8$ for a merger occurring at pericentre ($\nu_m=0$). Shown are choices for representative masses (colors) and projection angle (dashed lines).
}
    \label{fig:deltadelta_peri}
\end{figure}

Figure~\ref{fig:deltadelta_peri} shows how the tertiary pericentre distance $r_\mathrm{p, out}$ required to reach $\Delta\delta\mathrm{SNR}=8$ depends on tertiary mass, assuming $\nu_m=0$ (i.e., merger at pericentre) and a highly eccentric orbit. The different colours represent three choices for merging binary masses, appropriate for neutron star mergers, light BBHs and heavy BBHs. The relation is shown for two different choices of $\omega_0$. Note the characteristic scaling $r_\mathrm{p, out} \propto m_3^{1/2}$ that follows lines of constant acceleration. We also note how measurements of dephasing are in general more easily performed for lighter sources due to the increased number of GW cycles.

Taken together, Figures~\ref{fig:deltadelta} and \ref{fig:deltadelta_peri}, as well as additional analogous calculations, indicate that distinguishing eccentric-induced from circular-induced dephasing is feasible for nearby, low-mass binaries. The corresponding constraints on the triple are:

\begin{enumerate}
    \item \textbf{Tight eccentric triples:} The merging binary must be eccentric and have a pericentre distance that satisfies the results shown in Figure \ref{fig:deltadelta_peri}.
    
    \item \textbf{Merger near pericentre:} The merger should occur within approximately $-\frac{\pi}{2} < \nu_m < \frac{\pi}{2}$ for eccentricity effects to be distinguishable.
\end{enumerate}

\begin{figure}
    \centering
    \includegraphics[width=1\linewidth]{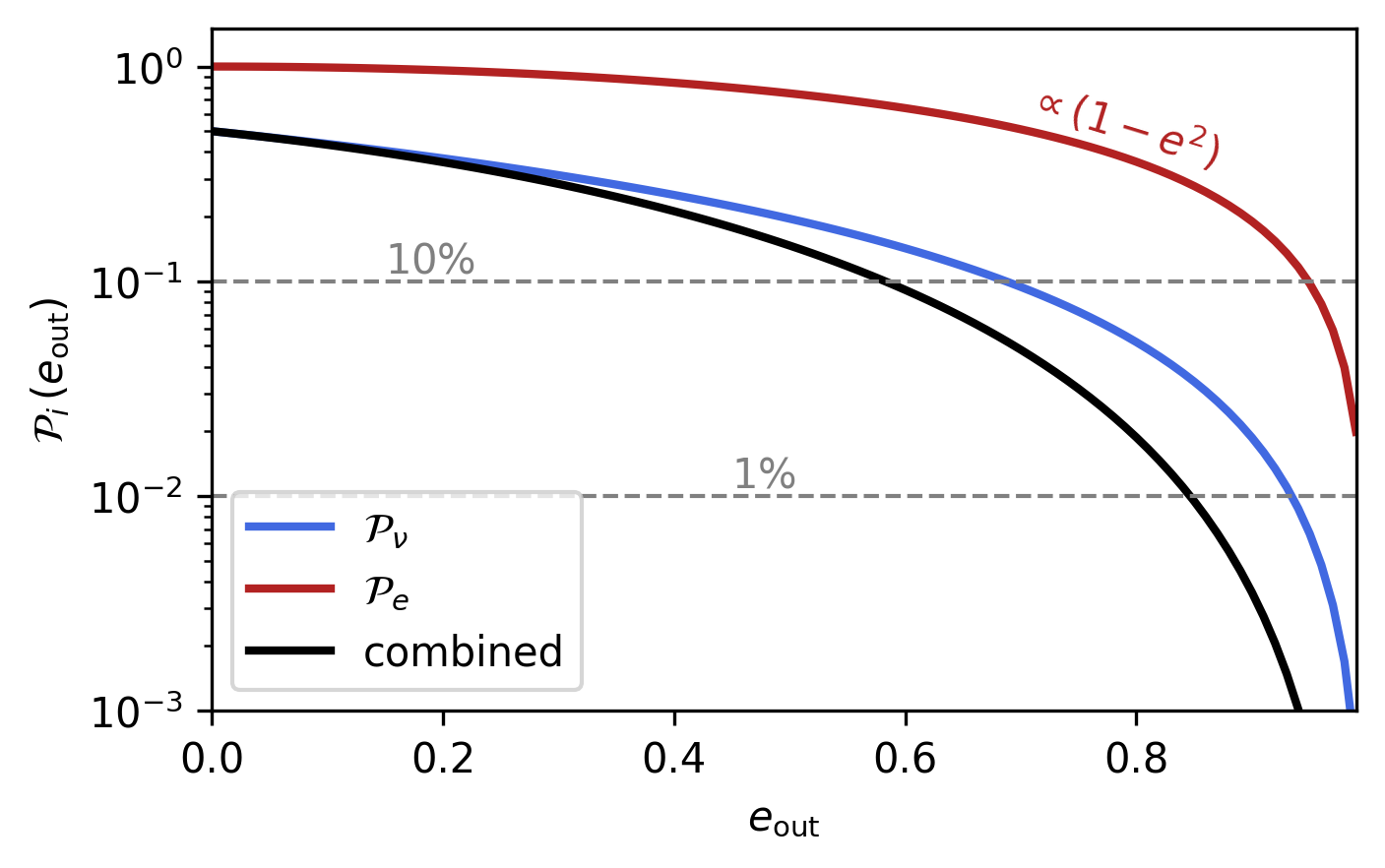}
    \caption{ Shown are the probability $\mathcal{P}_{\nu}$ of finding a binary with a true anomaly within $-\frac{\pi}{2}$ and $\frac{\pi}{2}$ as a function of eccentricty, as well as the probability $\mathcal{P}_e$ of finding a binary with $e> e_{\rm out}$, assuming a thermal distribution. These can be used to give an estimate of the fraction of binaries with the properties discussed in Sec. \ref{sec:exploration}.}
\label{fig:distr}
\end{figure}

\begin{figure*}
        \centering
        \begin{overpic}[width=.82\textwidth]{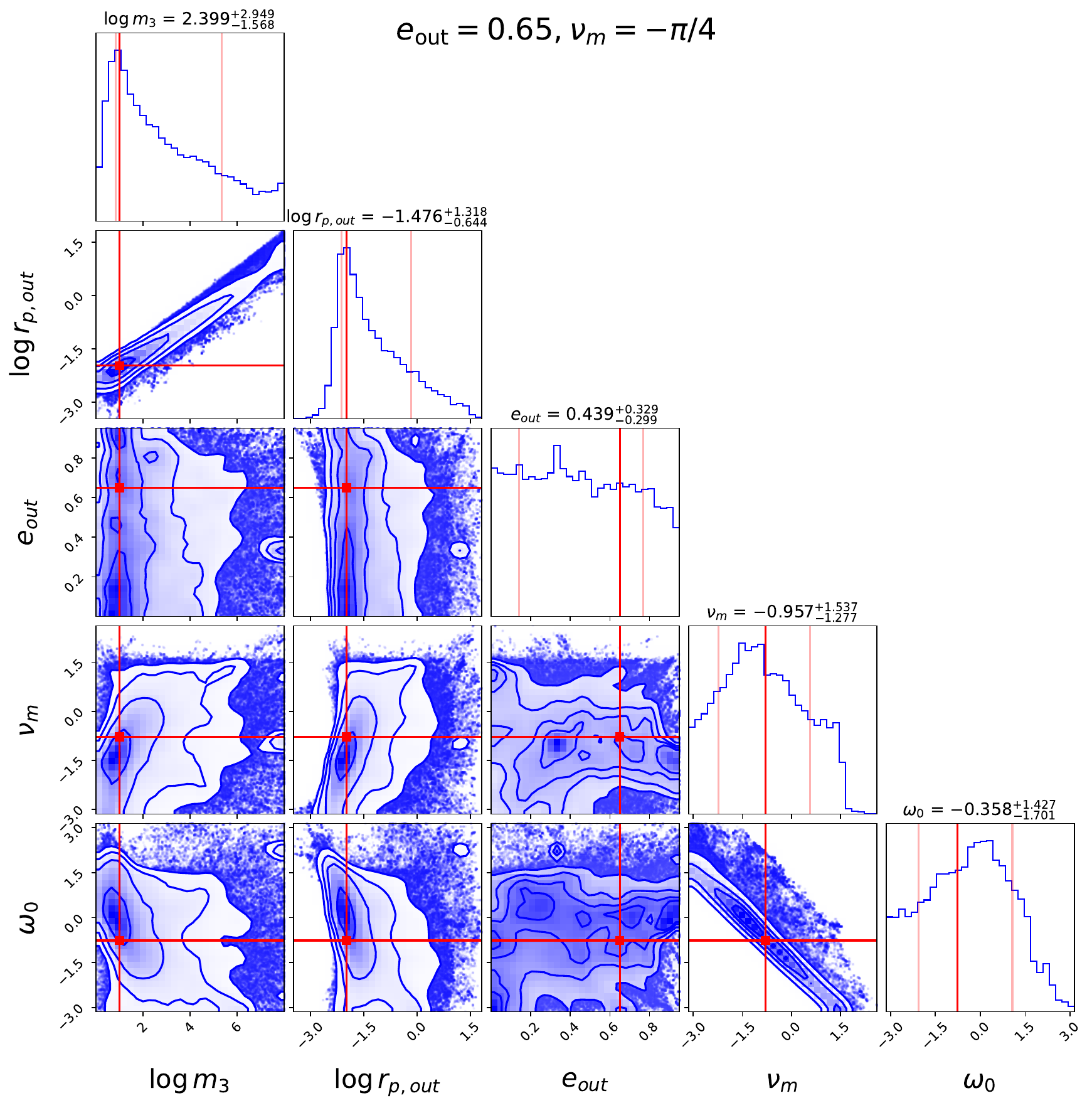}
            \put(65,65){\includegraphics[width=0.25\textwidth]{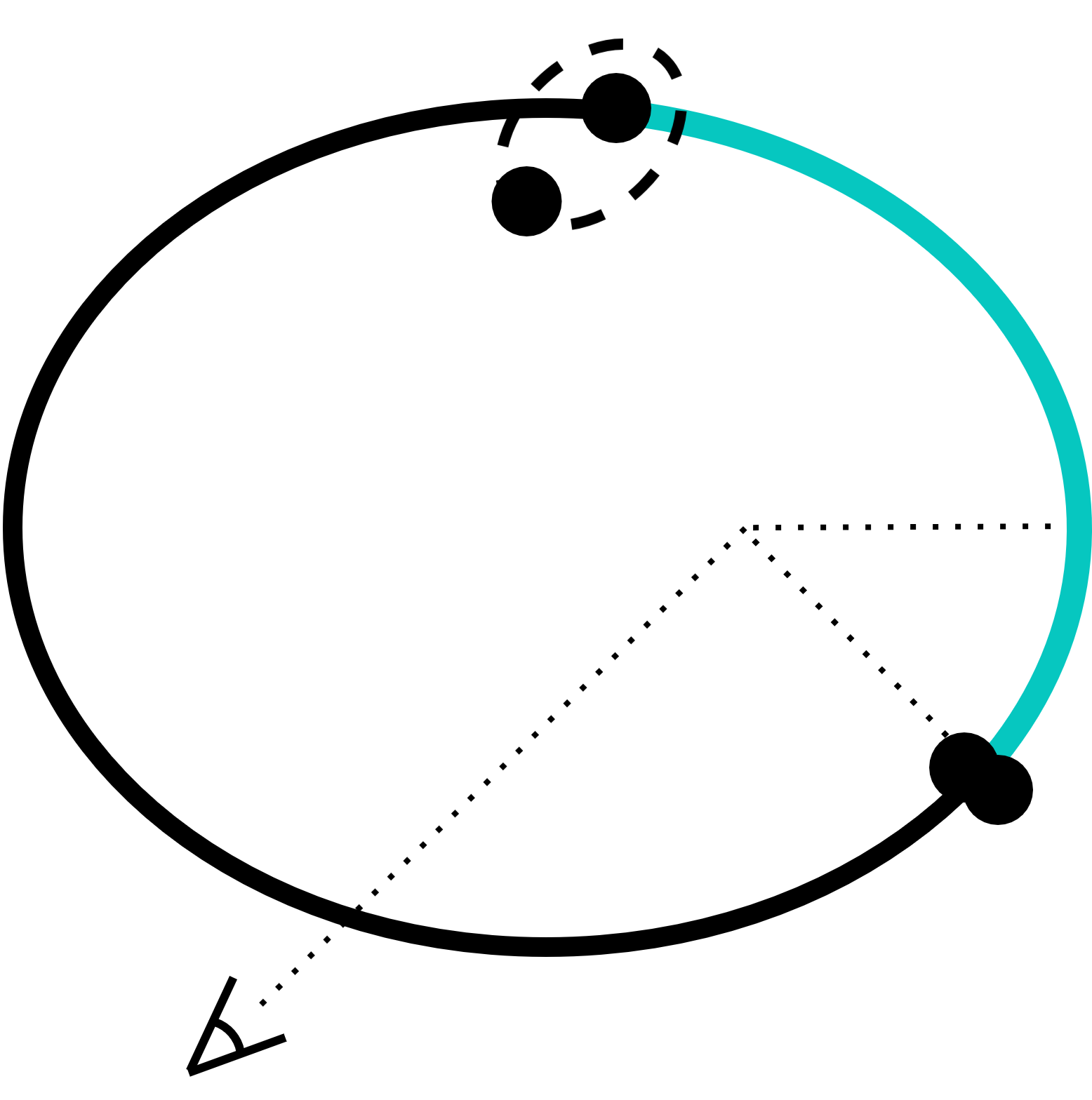}}
        \end{overpic}
        \caption{Corner plots for 5 outer orbital parameters $m_3, r_\mathrm{p, out}, e_\mathrm{out}, \nu_m,$ and $\omega_0$ for a three-body system observed by ET. The binary merges beyond pericentre, i.e. $\nu_m=-\pi/4$. We set $e_\mathrm{out}=0.65$ and $\omega_0 \sim -\pi/4$. We use $m_1 = 8M_\odot$, $m_2 = 8M_\odot$, $m_3 = 10M_\odot$, $z=0.2$, and $a_\mathrm{out}=0.03$AU. Above the panels, we show the recovered median ($50^{th}$ percentile) and bounds of the 68\% credible interval. The same bounds are visualised in the 1-d posteriors in shaded red, as well as the injected true values in bright red.}
        \label{fig:corner_e0.65}
\end{figure*}
    
\begin{figure*}
        \centering
        \begin{overpic}[width=.82\textwidth]{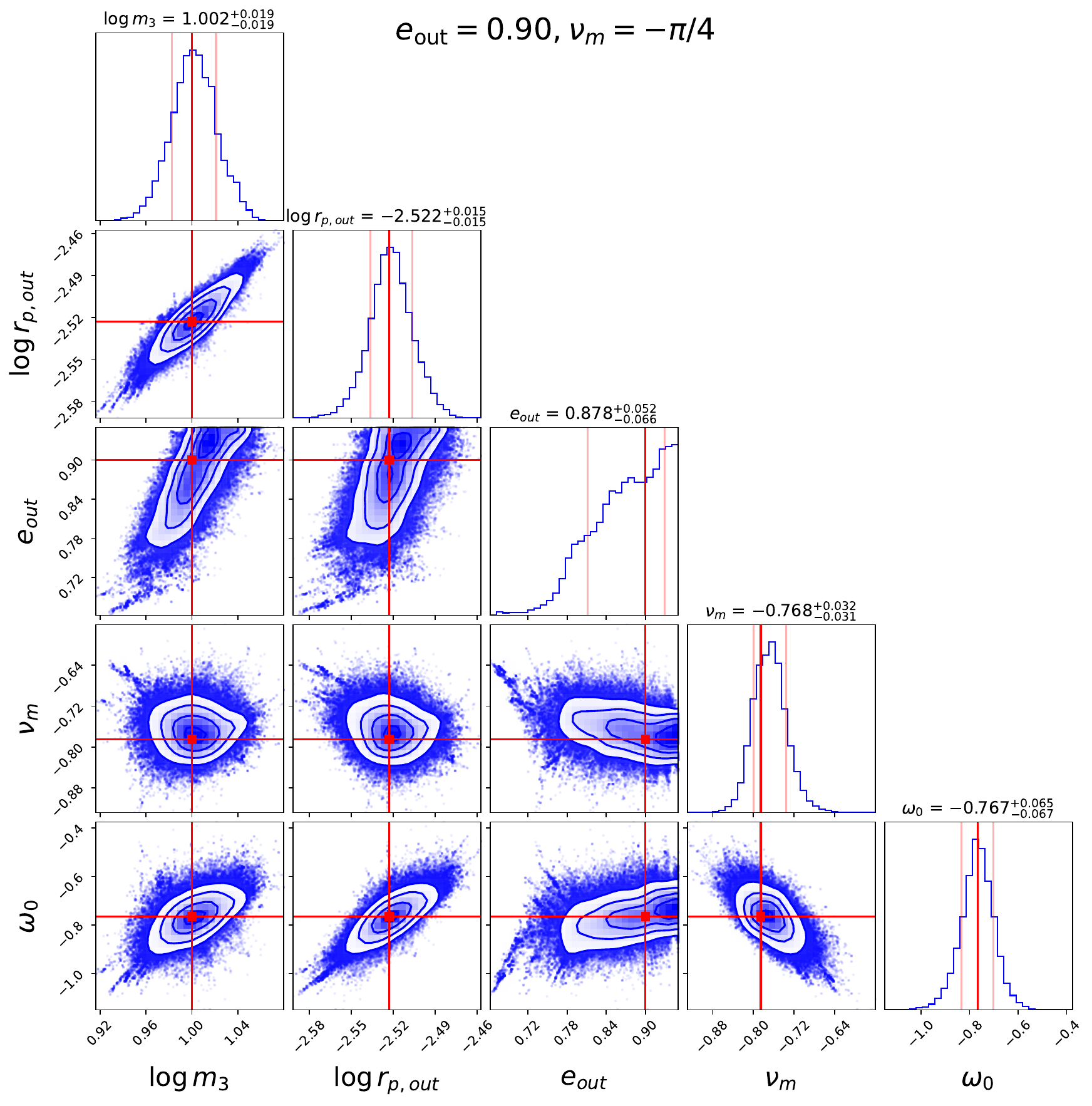}
            \put(65,65){\includegraphics[width=0.25\textwidth]{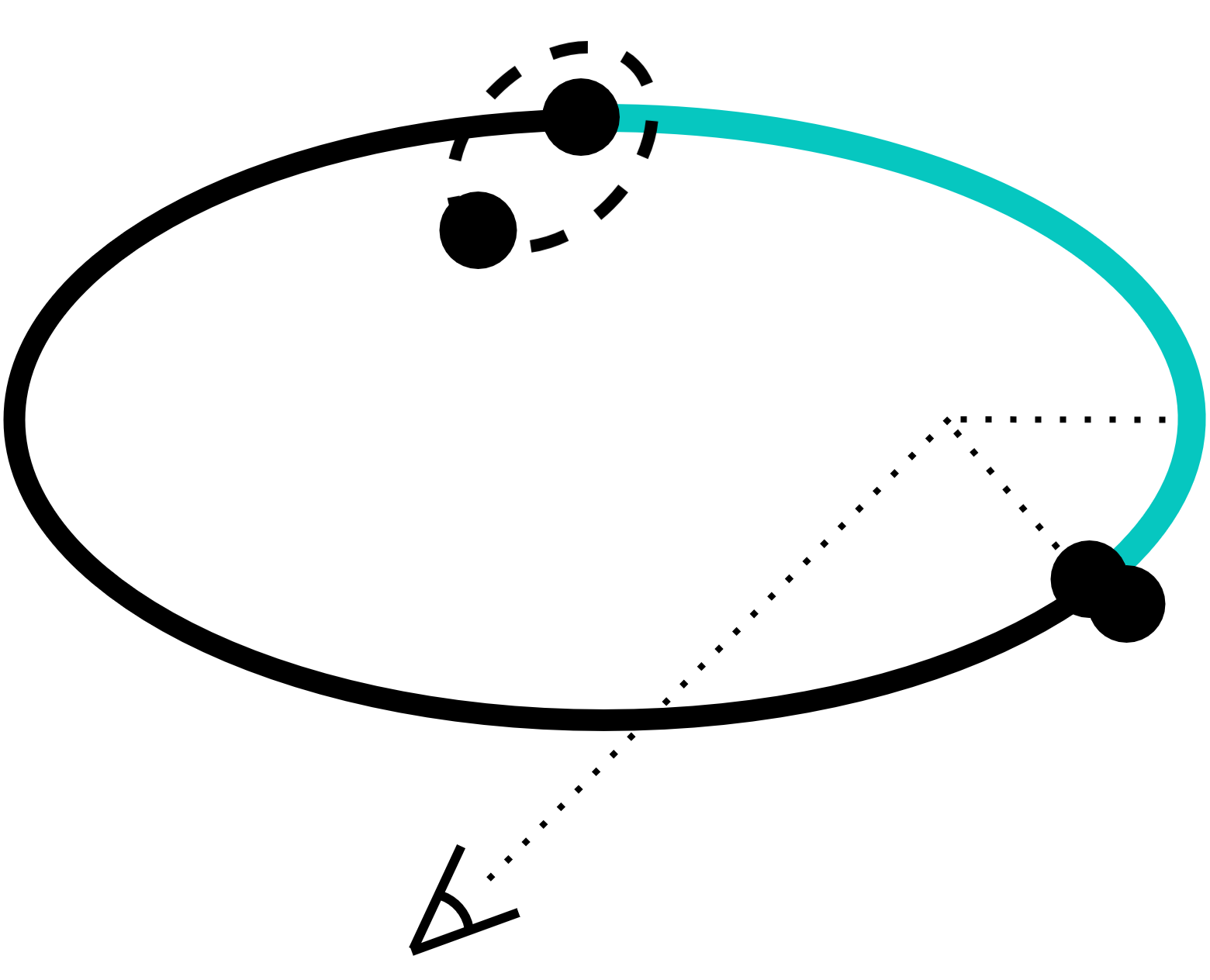}}
        \end{overpic}
        \caption{Corner plots for 5 outer orbital parameters $m_3, r_\mathrm{p, out}, e_\mathrm{out}, \nu_m,$ and $\omega_0$ for a three-body system observed by ET. The binary merges beyond pericentre, i.e. $\nu_m=-\pi/4$. We set $e_\mathrm{out} \sim 0.9$ and $\omega_0 \sim - \pi/4$. We use $m_1 = 8M_\odot$, $m_2 = 8M_\odot$, $m_3 = 10M_\odot$, $z=0.2$, and $a_\mathrm{out}=0.03$AU. Above the panels, we show the recovered median ($50^{th}$ percentile) and bounds of the 68\% credible interval. The same bounds are visualised in the 1-d posteriors in shaded red, as well as the injected true values in bright red.} 
        \label{fig:corner_e0.9}
\end{figure*}

We estimate the number of sources that satisfy these conditions, as well as other constraints. The binary masses used here correspond to the most common BBH mergers observed by LIGO, and we denote by $f_{\rm dyn}$ the fraction formed with a dynamically assembled tertiary. Focusing on the cluster channel, where the tertiary mass is comparable to the binary, detectable dephasing requires pericentre distances $\lesssim 10^{-2}\mathrm{AU}$. The overall probability combines the lognormal semimajor–axis distribution between $10^{-2}$ and $10^3$AU \citep{2016antonini}, a thermal eccentricity distribution, and the likelihood of observing a merger near pericentre. As shown in Fig. \ref{fig:distr}, for a representative semimajor axis of $0.1,\mathrm{AU}$ only a few percent of systems satisfy both $e_{\mathrm{out}} \gtrsim 0.8$ and $-\frac{\pi}{2} < \nu_m < \frac{\pi}{2}$. Since ET is expected to detect $10^5$–$10^6$ BBH mergers per year \citep[e.g.][]{2020maggioreet,2012PhRvD..86l2001R,2014PhRvD..89h4046R, 2019JCAP...08..015B}, of which a few percent lie at high SNR and low redshift, this still yields the possibility of a few to tens of events per year in the relevant part of parameter space after accounting for $f_{\rm dyn}$. The situation is more complex for the AGN channel, for which the eccentricity distribution of binaries around the SMBH is unknown, and most likely suppressed by gas effects. On the other hand, it is very likely that binaries merge in the vicinity of migration traps, which tend to be in a high acceleration region very close to the SMBH. Additionally, other than binaries around the SMBH, these traps might host stellar-mass three-body systems that are tight due to the surrounding gas, which may produce detectable phase shifts \citep{2025arXiv251115193T}.

It is important to emphasise that the $\mathcal{P}_\nu$ distribution shown in Fig. \ref{fig:distr} is a conservative estimate. While binaries on eccentric outer orbits spend most of their time near apocentre, where the orbital curvature and thus the induced acceleration are weakest, it is not a priori clear whether mergers predominantly occur in this regime. The chaotic nature of three-body interactions often triggers inspirals during close encounters, making mergers at small $\nu_m$ values, corresponding to the strong-curvature regime, entirely plausible \citep{samsing2018a, kai22024}. Furthermore, although the prior on the outer semi-major axis is taken to be uniform in $\log(a_\mathrm{out})$ over the range $10^{-2}-10^3$AU, the subset of systems that actually undergo mergers is preferentially drawn from the small-$a_\mathrm{out}$ end of this distribution. Such configurations naturally produce smaller pericentre distances, further enhancing the likelihood of mergers occurring in regimes where the induced dephasing is most pronounced.

\section{Parameter inference of mock events}
\label{sec:mock_inference}
We here report the results of the parameter inference with the MCMC framework described in Sec. \ref{sec:mcmc}. In this setup we keep the binary parameters, as well as the tertiary mass $m_3$ fixed: $m_1 = 8M_\odot, m_2 = 8 M_\odot, z=0.2, m_3 = 10M_\odot$. 

\subsection{Inference results}
\label{sec:inf_corner}

In Figs. \ref{fig:corner_e0.65} \& \ref{fig:corner_e0.9} we show corner plots for the outer orbit parameters of two MCMC runs. We show the results for a run with moderate eccentricity (0.65, Fig. \ref{fig:corner_e0.65}) and high eccentricity (0.9, Fig. \ref{fig:corner_e0.9}).


Fig. \ref{fig:corner_e0.65} corresponds to a moderately eccentric outer orbit on which the binary inspirals while passing the pericentre of the tertiary’s orbit. In this configuration, according to Eq. \ref{eq:losj} the projection effects dominate over the eccentricity contribution. The latter can only be dominant in the LOSJ approximation if $\nu_m = -\omega_0$. This signal has an SNR of 87 and is only marginally distinguishable from a vacuum waveform, with $\delta\mathrm{SNR} \sim 8.6$. Despite $\Delta \delta \mathrm{SNR} \sim 20$, meaning that it is clearly distinguishable from a generic LOSA dephasing (Eq. \ref{eq:dephasing_LOSA}), no tight constraints on the outer orbital parameters can be obtained as the LOSA (Eq. \ref{eq:losa-ecc}) does not change significantly over time. Note that this $\Delta \delta \mathrm{SNR}$ is in agreement with Fig. \ref{fig:deltadelta_peri}, where we see that this system, which has $p_\mathrm{out}\sim0.01$AU, lies just below the threshold of distinguishability for the red curve. This highlights that a large $\Delta\delta\mathrm{SNR}$ alone is insufficient; robust constraints on the outer–orbit parameters additionally require the signal to be measurably distinct from a vacuum waveform, i.e. to have a sufficiently high $\delta\mathrm{SNR}$. Importantly, the signal still carries enough power to reveal a clear degeneracy between $m_{3}$ and $r_{\mathrm{p,out}}$, as well as between $\nu_{m}$ and $\omega_{0}$, which become identical in the limit of a circular outer orbit. In this configuration, the time variation of the LOSA remains too limited to break these degeneracies. Even so, the data constrain $m_{3}$ to be below $\sim 10^{5}\,M_{\odot}$, thereby excluding a heavy supermassive black hole companion in this case.

In Fig. \ref{fig:corner_e0.9}, which corresponds to a highly eccentric outer orbit, the strong time variation of the line of sight acceleration produces a markedly more informative signal. Also here, the projection effects dominate. The event has a $\Delta\mathrm{SNR} \sim 48$ and a $\Delta\delta\mathrm{SNR} \sim 34.8$, and with $p_\mathrm{out} \sim 0.003$AU it is well below the detection threshold displayed in Fig. \ref{fig:deltadelta_peri}. The substantial dephasing relative to both the vacuum and generic LOSA waveforms results in significantly tighter constraints on the outer orbital parameters than Fig. \ref{fig:corner_e0.65}. Most notably, the degeneracy between $m_{3}$ and the outer pericentre distance $r_{\mathrm{p,out}}$, which is prominent at lower eccentricities, is effectively broken. The tertiary mass and $r_\mathrm{p, out}$ are now well constrained. Likewise, the degeneracy between $\nu_{m}$ and the projection angle $\omega_{0}$ is resolved, reflecting the fact that a highly eccentric configuration generates a rapidly evolving LOSA whose curvature carries distinct imprints of both quantities. Overall, the inference shows that for sufficiently eccentric outer orbits in tight 3-body systems the dynamics of the system are encoded strongly enough in the gravitational wave phase evolution that several of the relevant parameters can be recovered with high precision.

\subsection{Constraining the tertiary mass}
\begin{figure}[b]
    \centering
        \begin{overpic}[width=\linewidth]{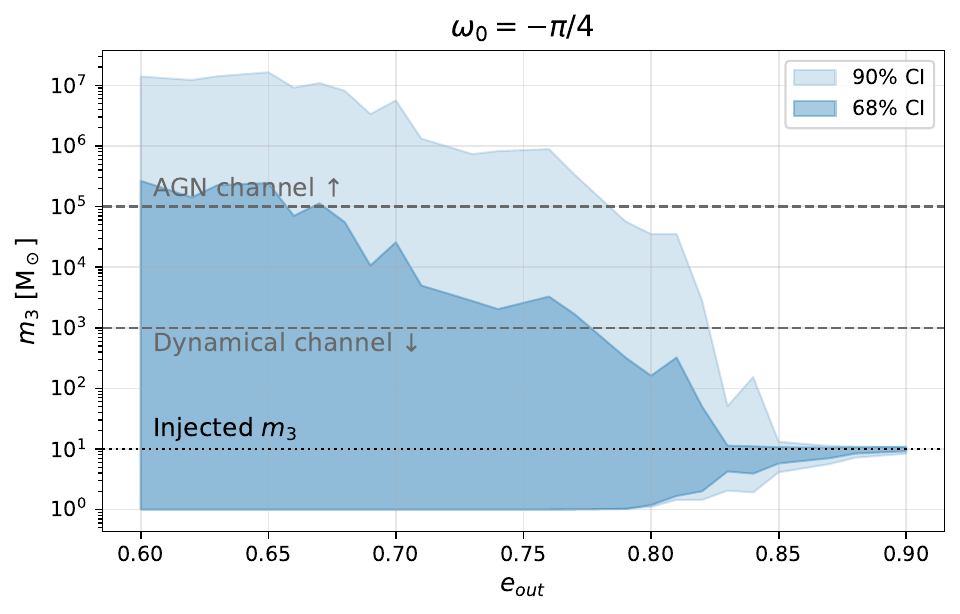}
            \put(76,26.5){\includegraphics[width=0.2\linewidth]{images/e0.9_fm-0.25pi.png}}
        \end{overpic}
        \begin{overpic}[width=\linewidth]{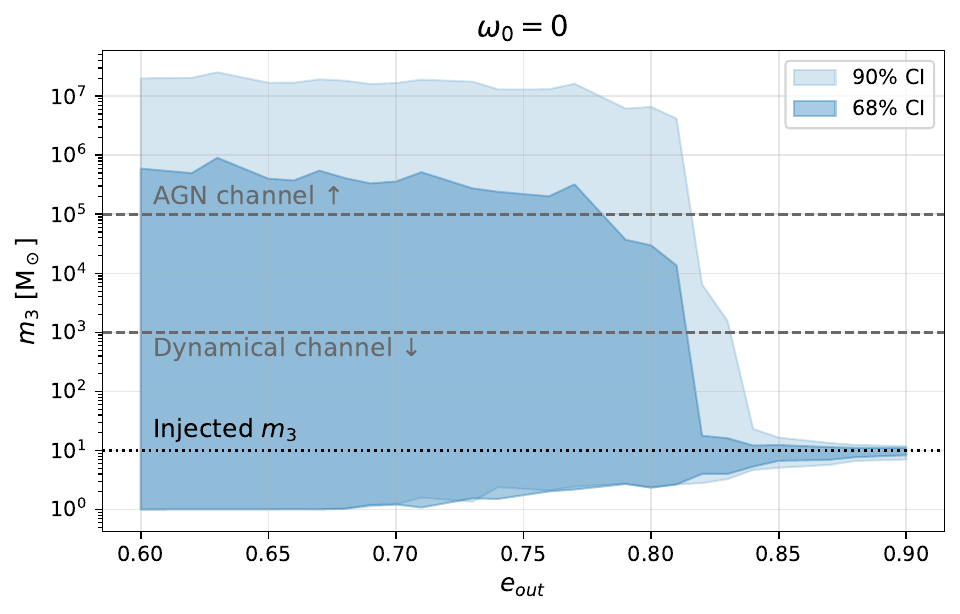}
            \put(72,30){\includegraphics[width=0.24\linewidth]{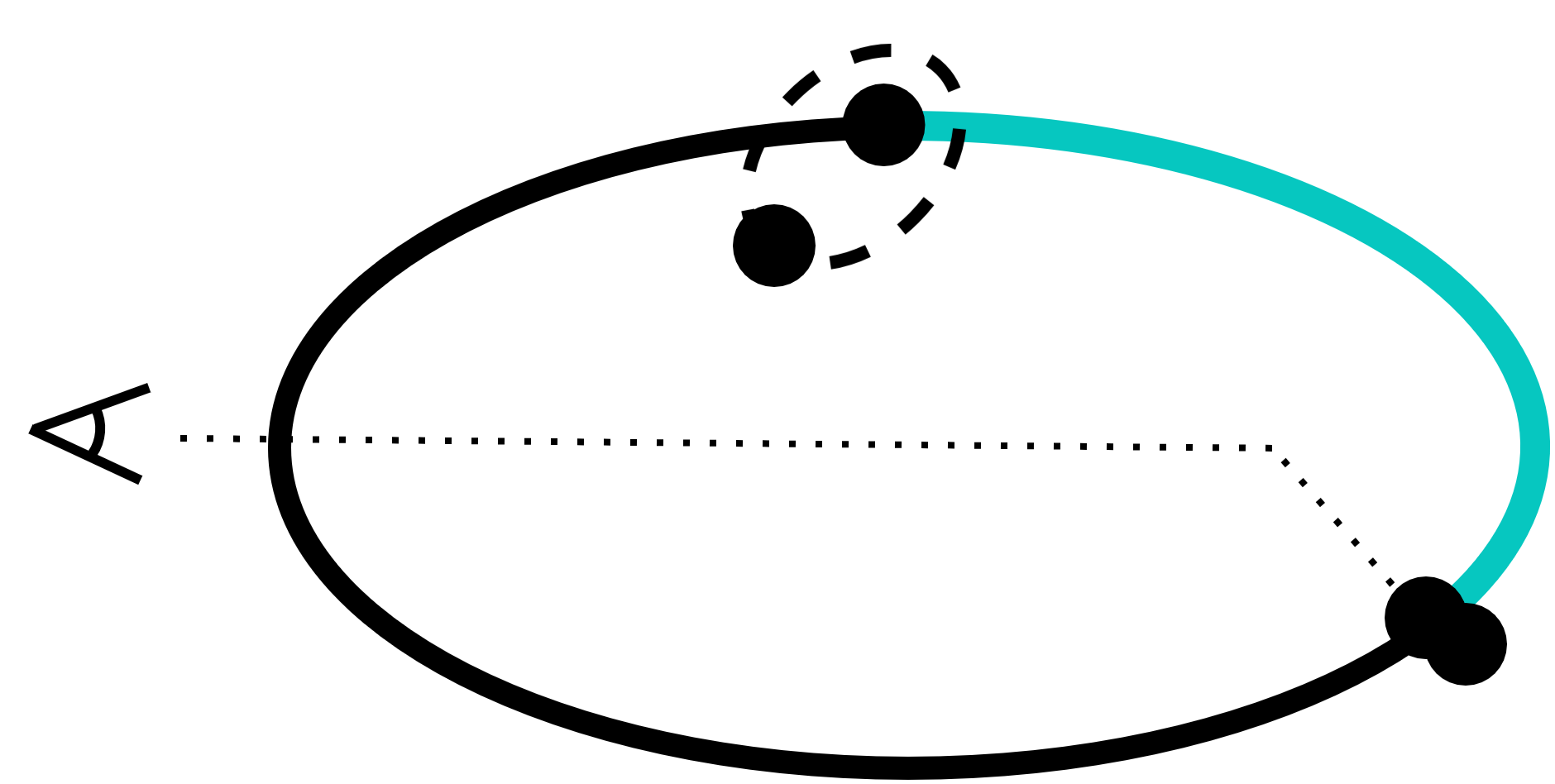}}
        \end{overpic}
    \vspace{-15pt}
    \caption{68\% and 90\% confidence intervals of the recovered tertiary mass as a function of outer eccentricity, for $\omega_0 = -\pi/4$ (top) and $\omega_0 = 0$ (bottom). We use $m_1=8M_\odot$, $m_2=8M_\odot$, $z=0.2$, $m_3 = 10M_\odot$, $a_\mathrm{out} = 0.04\mathrm{AU}$, and $\nu_m = -\pi/4$.}
    \label{fig:m3_eout_spread}
\end{figure}

\begin{figure}[t]
    \centering
    \begin{overpic}[width=\linewidth]{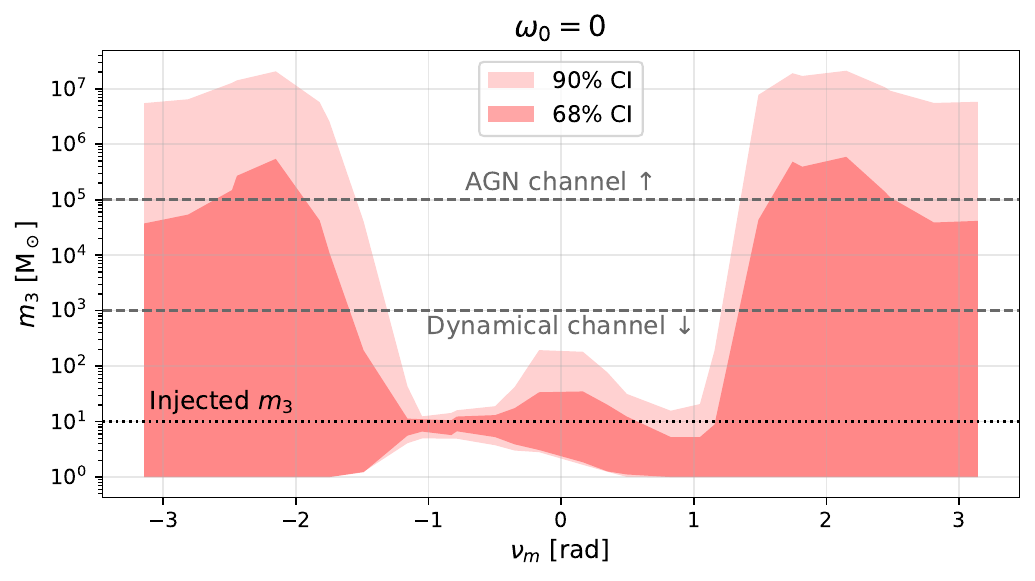}
            \put(72,16){\includegraphics[width=0.22\linewidth]{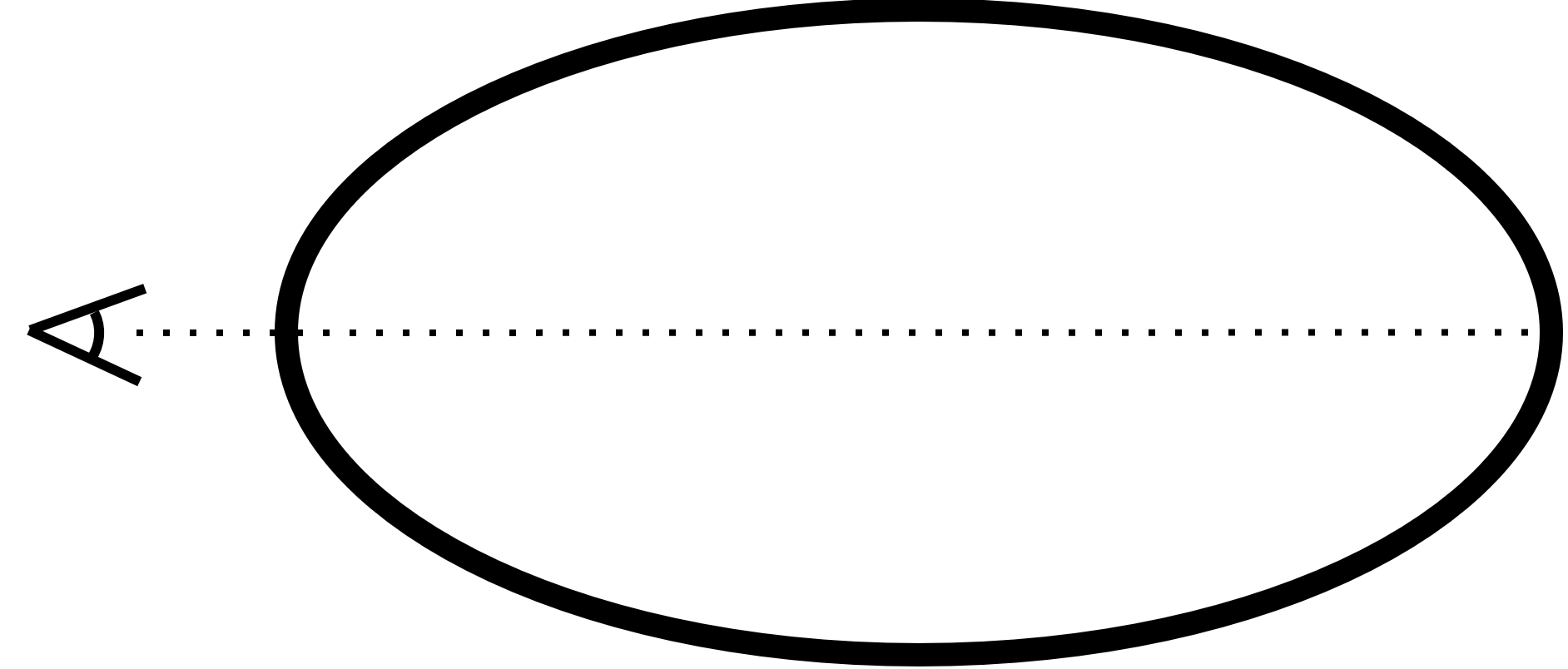}}
        \end{overpic}
        \begin{overpic}[width=\linewidth]{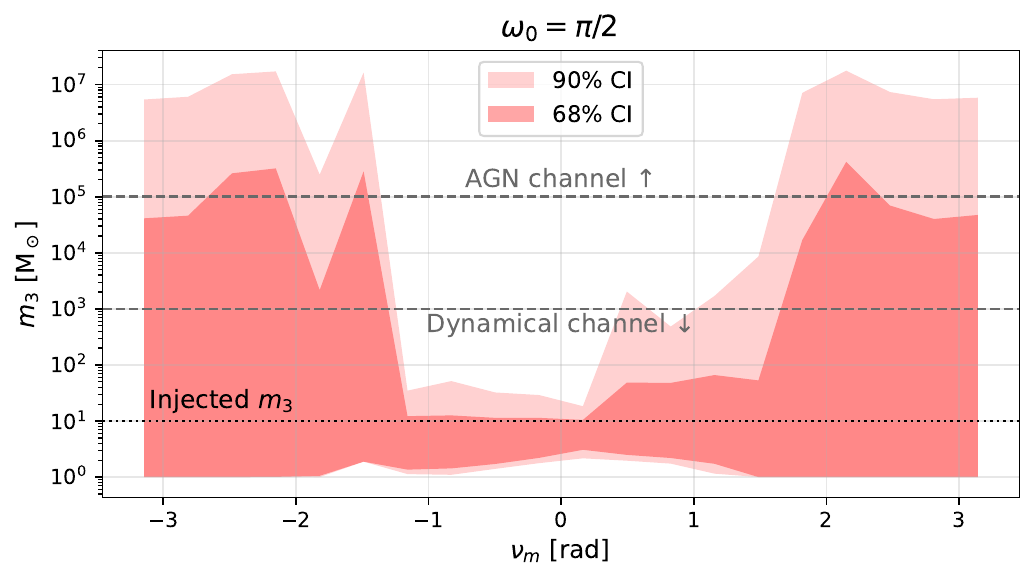}
            \put(76,16){\includegraphics[width=0.18\linewidth]{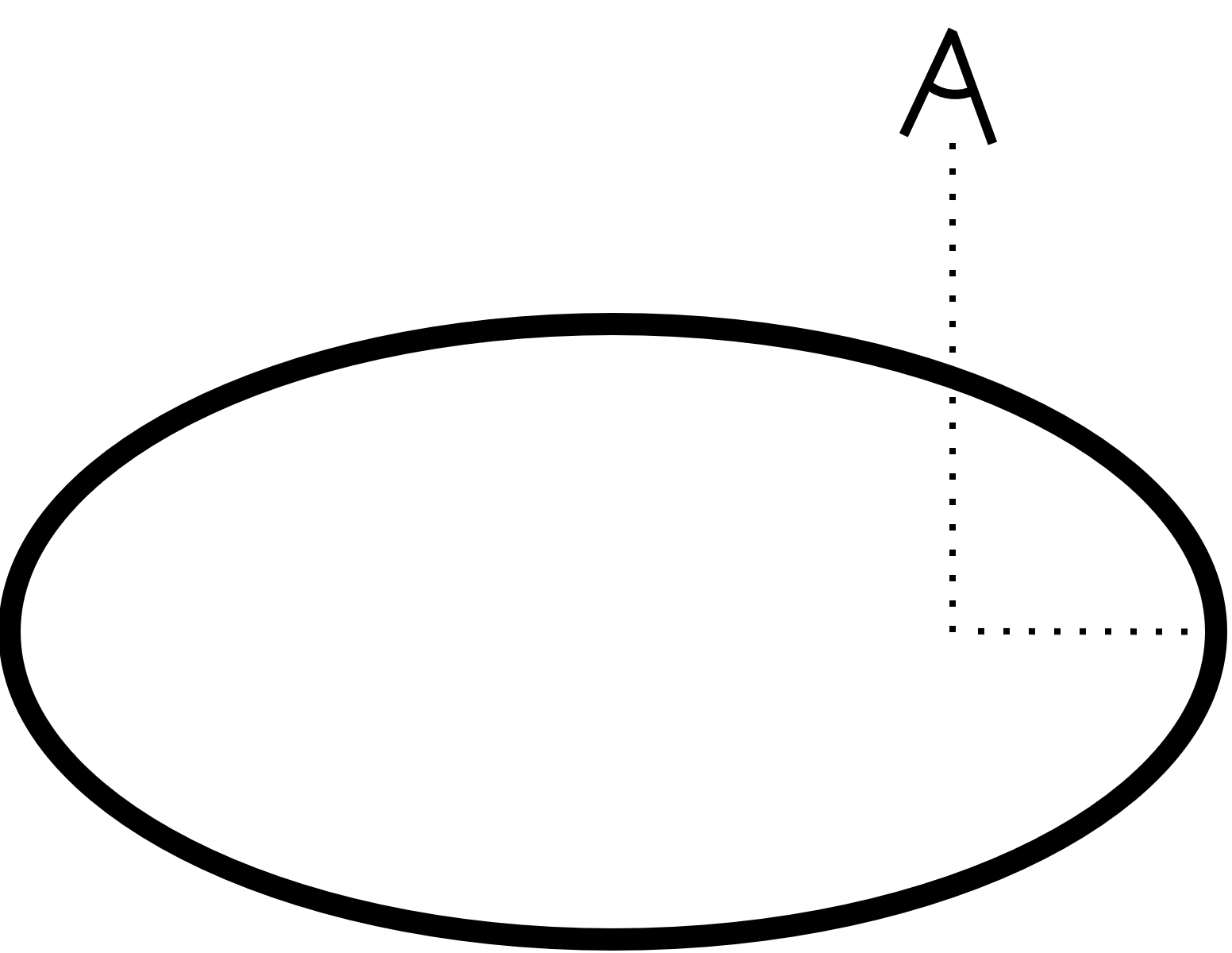}}
        \end{overpic}
    \vspace{-15pt}
    \caption{68\% and 90\% confidence intervals of the recovered tertiary mass as a function of true anomaly at merger $\nu_m$, for $\omega_0 = 0$ (top) and $\omega_0 = \pi/2$ (bottom). We use $m_1=8M_\odot$, $m_2=8M_\odot$, $z=0.2$, $m_3 = 10M_\odot$, $a_\mathrm{out} = 0.04\mathrm{AU}$, and $e_\mathrm{out} = 0.85$.}
    \label{fig:m3_fm_spread}
\end{figure}

In Fig. \ref{fig:m3_eout_spread} we show how the constraint on the tertiary mass depends on the outer eccentricity for a representative configuration, evaluated for two choices of the viewing orientation, $\omega_0 = 0$ and $\omega_0 = -\pi/4$. In both cases the binary merges at $\nu_m = -\pi/4$. A general trend is apparent: increasing outer eccentricity improves the mass constraint. For both orientations, the posterior on $m_3$ becomes constrained at the 68 percent level for $e_{\mathrm{out}} \gtrsim 0.82$ and at the 90 percent level for $e_{\mathrm{out}} \gtrsim 0.87$. At the 68 percent level the AGN channel can be excluded for $e_{\mathrm{out}} \approx 0.67$ for $\omega_0 = -\pi/4$ and for $e_{\mathrm{out}} \approx 0.78$ for $\omega_0 = 0$. A dynamical channel origin can be constrained at $e_{\mathrm{out}} \approx 0.73$ for $\omega_0 = -\pi/4$ and $e_{\mathrm{out}} \approx 0.82$ for $\omega_0 = 0$.

The two panels exhibit slightly different shapes. For $\omega_0 = 0$ a sharp change appears at high eccentricity that is not present in the top panel which shows a more gradual decrease in CI width. In the $\omega_0 = 0$ orientation, the outer orbit contains a point at $\nu = \pi/2$ where the acceleration vector becomes exactly perpendicular to the line of sight. As a result the projected acceleration drops to zero, which introduces a local decrease in the accumulated dephasing. This behaviour is a clear imprint of projection effects and therefore carries information that assists in constraining the tertiary mass. For eccentricities below approximately $e_{\mathrm{out}} \sim 0.8$ the binary either does not cross $\nu = \pi/2$ within the relevant segment of the orbit or it does so before entering the ET band. Once $e_{\mathrm{out}}$ exceeds this value the drop becomes observable in band, producing the corresponding improvement in the mass constraint. This feature underscores the need to model projection effects when extracting dephasing signatures from systems of this kind.

Especially visible in the lower panel, a flattening of the credible intervals toward lower eccentricities (with the 68 percent interval approaching $m_{3}\sim 10^{6}M_{\odot}$ and the 90 percent interval $\sim10^{7} M_{\odot}$) arises from the prior boundaries. Although we adopt broad priors on $m_{3}$ in the range $1$–$10^{8} M_{\odot}$, systems that are only weakly informative about the tertiary mass naturally accumulate against these limits, producing the apparent saturation in the intervals.

The overall morphology of the credible-interval curves in Fig. \ref{fig:m3_fm_spread} closely mirrors that seen in Fig. \ref{fig:deltadelta}. For $\omega_0 = 0$ the most informative mergers occur for $\nu_m \sim \pm 1$ rad, offset from pericentre. For $\omega_0 = \pi/2$, mergers closest to pericentre provide the strongest constraint, illustrating that the optimal phase for inference depends on the viewing geometry. Perhaps unexpectedly, Fig. \ref{fig:m3_fm_spread} also shows that the mass constraint is not significantly degraded at apocentre. Instead, the weakest constraints arise somewhat before and after this point ($\nu_m \sim \pm 2$ rad). This behaviour has a simple underlying explanation: when the binary inspirals near apocentre, the centre of mass undergoes a noticeable change in projected direction, introducing curvature in the line-of-sight acceleration. This time-dependent modulation of the projection factor $F(\omega_0, t)$ provides additional structure in the waveform, allowing the LOSA signature to be more cleanly disentangled and yielding a slightly improved constraint on $m_3$ compared to configurations immediately adjacent to apocentre. We note that this behaviour differs from that seen in Fig. \ref{fig:deltadelta}, where apocentre configurations produce the poorest constraints. This comparison highlights that individual parameter constraints can exhibit subtleties that are not captured by a global metric such as the $\delta \Delta \mathrm{SNR}$, which effectively marginalises over the full parameter space.

We also note a distinctive feature in the bottom panel in the form of a dip around $\nu_m \simeq -1.8$ rad. This point is noteworthy because it corresponds to a configuration in which the binary enters the detector band precisely as the LOSA becomes perpendicular to the observer, producing the characteristic dip in the accumulated dephasing at the moment when its amplitude is largest. This is the same feature causing the drop in the bottom panel of Fig. \ref{fig:m3_eout_spread}. For $\nu_m$ values lower than $\sim -1.8$ rad, the dip lies outside the observing band, whereas for $\nu_m$ slightly higher than this value, the dip shifts to higher frequencies where the dephasing amplitude is smaller, making the effect less pronounced and thereby yielding weaker constraints on $m_3$. At higher resolution the feature would appear smoother, but its origin and behaviour remain clear.

As established in Sec. \ref{sec:exploration}, the regime $-\frac{\pi}{2}\leq \nu_m \leq \frac{\pi}{2}$ (and, for the present configuration, more precisely within $\pm 1$ rad) is where the waveform departs most strongly from a generic LOSA-type dephasing and where the tightest constraints on $m_3$ are expected. Importantly, however, valuable astrophysical insights can be gained without a full measurement of the tertiary mass. Even the ability to place meaningful upper limits on $m_3$ is sufficient to rule out or favour specific formation channels, and such constraints can be obtained for systems that are less tightly bound than the fiducial configuration. For example, to constrain the dynamical channel at the $68\%$ level, a merger occurring near pericentre may have an outer semimajor axis larger by a factor of $\sim 1.5$ than $a_{\mathrm{out}} = 0.04\,\mathrm{AU}$ for $\omega_0 = 0$, and by $\sim 2$ for $\omega_0 = \pi/2$. Similarly, to exclude the AGN channel, $a_{\mathrm{out}}$ can be increased by a factor of $\sim 3$ for $\omega_0 = 0$ or by $\sim 2.75$ for $\omega_0 = \pi/2$. These scalings demonstrate that robust constraints on formation pathways can be achieved even when the system is not in the most informative region of parameter space.

\section{Analysis of GW190814 and O4a events}
\label{sec:lvk_inference}

\begin{figure*}[t]
    \centering
    \includegraphics[width=\linewidth]{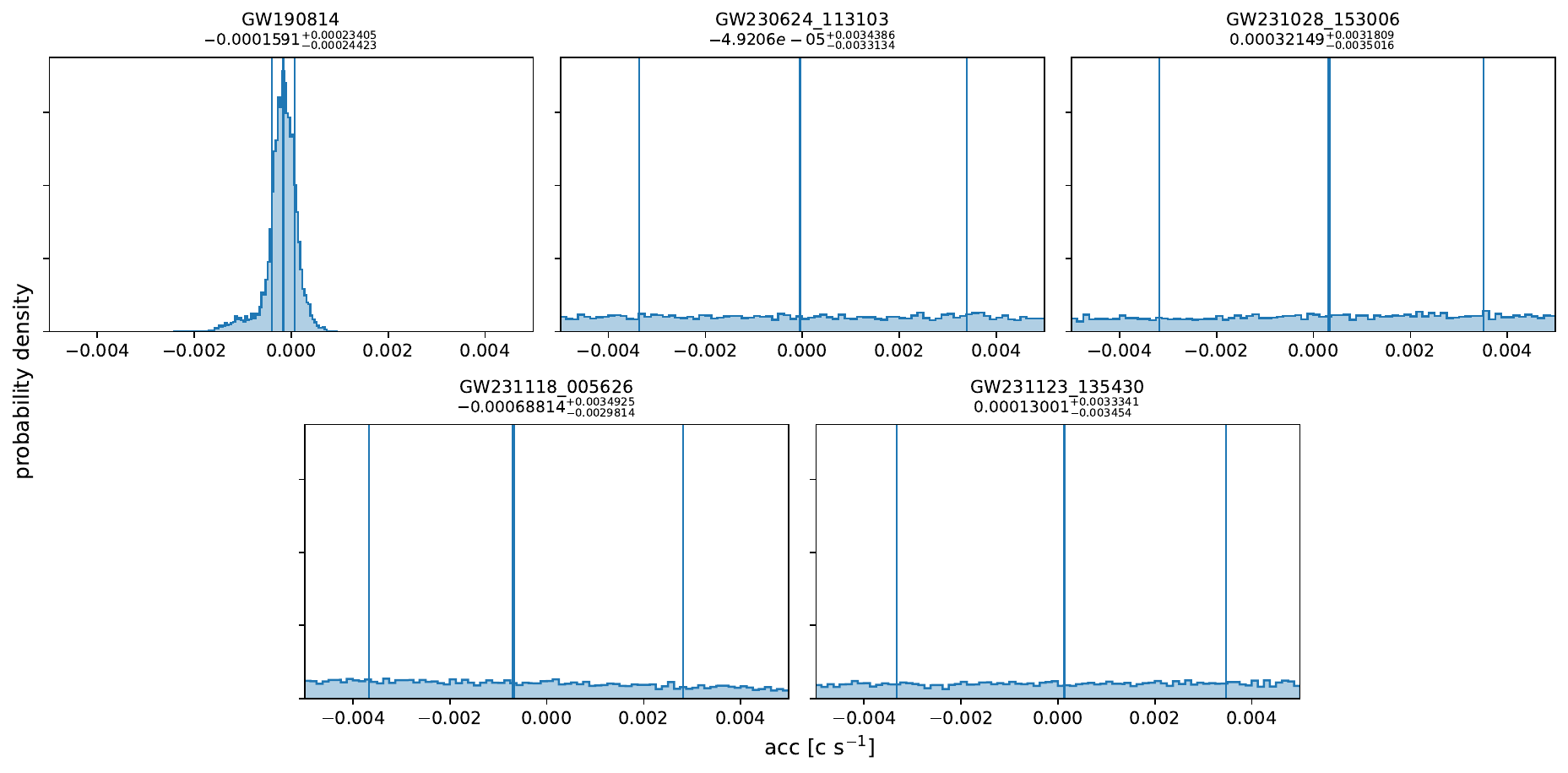}
    \caption{Distributions for the inferred LOSA from 5 different GW events. For GW190814 we find a constraint consistent with zero LOSA, while the remaining four events are unconstrained and return the prior distribution. For each panel, the quoted value corresponds to the 50th percentile of the posterior, with uncertainties given by the 16th and 84th percentiles. These percentiles are also indicated within the panels, defining the central 68\% credible interval.}
    \label{fig:LOSA-posteriors}
\end{figure*}

In this section we discuss the results from our LOSA analysis of several LVK events, most notably GW190814 \citep{abbottGW190814}. Recent work by \cite{2024Han} reported a LOSA of $0.0015^{+0.0008}_{-0.0008} c s^{-1}$ for this particular event, suggesting the presence of a nearby compact object. Additionally, we analyse 4 events in the LVK O4a catalogue \citep{gwtc-4} which contained reported discrepancies between different waveform models, potentially due to an unmodelled LOSA. For all events, we initially only fit the generic LOSA of Eq. \ref{eq:dephasing_LOSA}. Only those events for which we are able to constrain a non-zero acceleration, we would re-analyse with our eccentric dephasing model. We use wide priors for the O4a observations and adopt the GW190814 priors from \cite{2024Han}. For all analyses we assume a uniform prior distribution for the acceleration between $-0.005 c s^{-1}$ and $+0.005 c s^{-1}$. The noise power spectral densities were obtained for each individual event directly from the GWTC (\url{https://gwosc.org/}).

The results are depicted in Fig. \ref{fig:LOSA-posteriors}. For the 4 O4a events, our posteriors return the prior distribution. Therefore, we are not able to constrain any LOSA. The dephasing is largest and therefore most easily constrained early in the inspiral at low frequencies (Eq. \ref{eq:dephasing_LOSA}). Binaries with low component masses and small mass ratios (non-equal masses) spend more time in this regime and are therefore better candidates for a detection. The 4 LVK events have neither of these properties. With this finding we are able to rule out that the inference discrepancies between different waveform models for these events are caused by a missing LOSA.

For GW190814, the inferred posterior is consistent with no LOSA, with 0 lying within 1-$\sigma$ of the mean. Importantly, this result is in line with the non-detection reported in \cite{2025arXiv250622272T} and contradicts the claimed LOSA observation presented in \cite{2024Han}. In reanalysing the data presented in \cite{2024Han} using identical parameter settings, we successfully reproduce the LOSA reported therein. However, we find that their use of a 4-second signal duration, corresponding to an effective inspiral phase of only 1.6 seconds, is insufficient for analysing a signal of roughly 10 seconds in total inspiral duration \citep{abbottGW190814}. When repeating their analysis with a more appropriate and conservative 32-second segment, we find no evidence for any LOSA in GW190814. Our results therefore demonstrate that the inference reported in \cite{2024Han} arises from their limited signal window rather than from any genuine physical effect.

\section{Summary and conclusion}
We summarise the main findings of this work below.

\begin{itemize}
    \item This work introduces a dephasing model required for dynamical three-body interactions that self-consistently treats the full outer orbit, allowing for arbitrary eccentricity and projection effects. By capturing the resulting time-dependent LOSA, the model breaks the fundamental mass–distance degeneracy inherent to constant-acceleration prescriptions and enables constraints on the tertiary mass and orbital parameters.
    \item Two conditions in the three-body configuration apply in order for eccentric dephasing with projection effects to be distinguishable from generic LOSA dephasing ($\Delta \delta \mathrm{SNR} \gtrsim 8$) with ET sensitivity. Firstly, the outer orbit must be at least appreciably eccentric ($e_{\mathrm{out}} \gtrsim 0.7$) and sufficiently tight, with a semi-major axis of order $0.01$ to $0.1\,\mathrm{AU}$ for stellar–mass tertiaries. Secondly, the merger must occur close to pericentre, i.e. approximately within the interval $-\frac{\pi}{2} < \nu_m < \frac{\pi}{2}$. In this regime, the LOSA varies most strongly and the eccentric signature is maximally imprinted. Apart from a sufficient $\Delta\delta \mathrm{SNR}$, a detectable eccentric dephasing additionally needs enough power in its $\delta \mathrm{SNR}$ ($\gtrsim 8$) for it to be distinguishable from a vacuum waveform.
    \item The required ingredients for observable eccentric dephasing and constraining the outer orbital parameters are astrophysically plausible. Under simple assumptions about the semimajor axis and eccentricity distributions of triples in clusters, a few percent of dynamically formed BBH triples satisfy the combined requirements of high outer eccentricity and merger near pericentre. When this fraction is folded together with ET detection forecasts, we still expect a few to tens of events per year in which eccentric Rømer dephasing should be observable, depending on the poorly known fraction $f_{\mathrm{dyn}}$ of BBHs assembled in three-body systems.
    \item For N-body simulations of three-body interactions in dense stellar clusters that produce measurable phase shifts, the relevant parameters to record are the outer true anomaly at merger $\nu_m$, eccentricity $e_{\mathrm{out}}$, and semi-major axis $a_{\mathrm{out}}$. Characterising the distributions of these quantities will give more accurate predictions for their detection rates and substantially improve our understanding of the dynamical assembly of BHs in three-body systems.
    \item Across much of the parameter space the tertiary mass cannot be measured precisely but its upper limit can still be tightly constrained, providing strong discriminating power between dynamical and AGN channels. For the slightly optimised but astrophysically plausible configurations explored here, the mass becomes measurable once the outer eccentricity exceeds $e_\mathrm{out} \sim 0.8$ with $a_\mathrm{out} = 0.04$AU (i.e. $r_\mathrm{p, out} \sim 0.008$AU), with the exact threshold depending on the merger true anomaly and the viewing angle. Even at moderate eccentricities ($e_\mathrm{out}< 0.8$), for mergers near pericentre and for small outer separations, as well as mergers with high $e_\mathrm{out}$ outside the optimal $\nu_m$ range, the recovered upper limit on $m_3$ can be sufficient to rule out supermassive perturbers in the AGN channel.
    \item Reanalysis of GW190814 and several O4a events finds no evidence for a LOSA. The previously claimed evidence in GW190814 \citep{2024Han} is not reproduced when an appropriate signal duration is used, confirming that the effect is an artefact of an insufficiently long data segment. Additionally, we can rule out that the waveform discrepancies reported in the selected O4a events are caused by an unmodelled LOSA as their inference returns the (flat) prior distributions.
\end{itemize}



This work opens several avenues for further investigation. A key next step is to incorporate the eccentricity of the inner binary, which is expected to significantly enhance the detectability of Rømer dephasing through the additional harmonic structure of eccentric inspirals \citep{2025arXiv251104540Z, pedpaper}. This extension will also allow direct application of the model to LVK events with confirmed non-zero eccentricity. On the astrophysical side, more detailed predictions for the distribution of triple configurations can be obtained through dedicated N-body simulations of dense stellar clusters and, potentially, of three-body interactions in AGN disks. Finally, while our analysis has focused on stellar-mass triples, an analogous injection study with supermassive tertiary companions would shed light on the astrophysical ingredients for detectable dephasing in the AGN channel and the need for using full orbital models. Together, these developments will help build a more complete framework for identifying and directly characterising EEs in GW sources.

\section*{Data availability}

The data used to complete this study will be shared upon request to the corresponding author.

\section*{Acknowledgements}
The authors are thankful for insightful discussions with Rico K. L. Lo, which have been extremely helpful in analysing the LVK events. K.H., P.S. and J.S. are supported by the Villum Fonden grant No. 29466, and by the ERC Starting Grant no. 101043143 – BlackHoleMergs. L.Z. is supported by the European Union’s Horizon 2024 research and innovation program under the Marie Sklodowska-Curie grant agreement No. 101208914. J.T. is supported by the Alexander von Humboldt Foundation under the project no. 1240213 - HFST-P. The Center of Gravity is a Center of Excellence funded by the Danish National Research Foundation under grant No. 184. The Tycho supercomputer hosted at the SCIENCE HPC center at the University of Copenhagen was used for performing the parameter inference of the LVK events presented in this work. The authors are grateful for their Jupyter notebook epic\_ecc\_PE\_MCMC.ipynb.

\newpage
\newpage
\bibliography{apssamp}

\end{document}